\documentclass[10pt]{article}

\usepackage{adjustbox}

\usepackage{lineno}

\usepackage{color}
\usepackage[normalem]{ulem}

\usepackage{amsmath,latexsym,color,subfigure,setspace,multirow,soul, xspace}
\usepackage{url}\RequirePackage[colorlinks,citecolor=blue, linkcolor=black,urlcolor = black]{hyperref}

\usepackage{amssymb}

\usepackage{graphicx}
\usepackage{epstopdf}
\usepackage{pdflscape}
\usepackage{caption}
\usepackage{subfigure}

\usepackage[authoryear,round,semicolon,sort]{natbib} 

\usepackage{xr}

\usepackage[labelfont=bf,labelsep=period,justification=justified]{caption}

\makeatletter
\renewcommand{\@biblabel}[1]{\quad#1.}
\makeatother

\usepackage{lscape}
\usepackage{adjustbox}
\usepackage{array}
\usepackage{relsize}
\usepackage{sidecap}
\usepackage{bm}
\usepackage{bbm}
\usepackage{multicol}
\usepackage{float}
\usepackage{booktabs}
\usepackage{multirow,multicol}
\usepackage[ruled,vlined]{algorithm2e}

\def\eq#1{(\ref{#1})}

\def\beginmat{ \left( \begin{array} }
\def\endmat{ \end{array} \right) }

\def\log{{\rm log}}

\def\cond{\, | \,}
\newcommand*\diff{\mathop{}\!\mathrm{d}}
\newcolumntype{P}[1]{>{\centering\arraybackslash}p{#1}}

\date{}

\newcommand{\bg}{\mathbf{g}}

\newcommand{\bx}{\mathbf{x}}
\newcommand{\by}{\mathbf{y}}

\newcommand{\bw}{\mathbf{w}}

\newcommand{\bK}{\mathbf{K}}
\newcommand{\bV}{\mathbf{V}}
\newcommand{\bA}{\mathbf{A}}
\newcommand{\bB}{\mathbf{B}}
\newcommand{\bC}{\mathbf{C}}
\newcommand{\bX}{\mathbf{X}}

\newcommand{\bH}{\mathbf{H}}

\newcommand{\bW}{\mathbf{W}}

\newcommand{\bZ}{\mathbf{Z}}
\newcommand{\bD}{\mathbf{D}}
\newcommand{\bI}{\mathbf{I}}

\newcommand{\T}{\intercal}
\newcommand{\wt}{\widetilde}
\newcommand{\wh}{\widehat}

\newcommand{\E}{\mathbb{E}}
\newcommand{\V}{\mathbb{V}}

\newcommand{\N}{\mathcal{N}}

\newcommand{\bvarepsilon}{\boldsymbol\varepsilon}
\newcommand{\bbeta}{\boldsymbol\beta}

\newcommand{\bgamma}{\boldsymbol\gamma}
\newcommand{\bdelta}{\boldsymbol\delta}

\newcommand{\bvartheta}{\boldsymbol\vartheta}

\newcommand{\bphi}{\boldsymbol\phi}

\newcommand{\bxi}{\boldsymbol\xi}

\newcommand{\bomega}{\boldsymbol\omega}
\newcommand{\btau}{\boldsymbol\tau}

\newcommand{\bXi}{\boldsymbol\Xi}

\newcommand{\bSigma}{\boldsymbol\Sigma}

\usepackage{etoolbox} 
\makeatletter 
\newcommand*\linenomathpatch{\@ifstar{\linenomathpatch@AMS}{\linenomathpatch@}}
\newcommand*\linenomathpatch@[1]{
  \expandafter\pretocmd\csname #1\endcsname {\linenomathWithnumbers}{}{}
  \expandafter\pretocmd\csname #1*\endcsname{\linenomathWithnumbers}{}{}
  \expandafter\apptocmd\csname end#1\endcsname {\endlinenomath}{}{}
  \expandafter\apptocmd\csname end#1*\endcsname{\endlinenomath}{}{}
}
\newcommand*\linenomathpatch@AMS[1]{
  \expandafter\pretocmd\csname #1\endcsname {\linenomathWithnumbersAMS}{}{}
  \expandafter\pretocmd\csname #1*\endcsname{\linenomathWithnumbersAMS}{}{}
  \expandafter\apptocmd\csname end#1\endcsname {\endlinenomath}{}{}
  \expandafter\apptocmd\csname end#1*\endcsname{\endlinenomath}{}{}
}
\let\linenomathWithnumbersAMS\linenomathWithnumbers
\patchcmd\linenomathWithnumbersAMS{\advance\postdisplaypenalty\linenopenalty}{}{}{}
\makeatother 

\linenomathpatch{equation}
\linenomathpatch*{gather}
\linenomathpatch*{multline}
\linenomathpatch*{align}
\linenomathpatch*{alignat}
\linenomathpatch*{flalign}

\newcommand{\red}{\textcolor{black}}

\newcolumntype{R}[2]{%
    >{\adjustbox{angle=#1,lap=\width-(#2)}\bgroup}%
    l%
    <{\egroup}%
}

\allowdisplaybreaks

\begin{document}

\begin{flushleft}
{\Large
\textbf{A Simple Approach for Local and Global Variable Importance in Nonlinear Regression Models}
}
\newline
\\
Emily T.~Winn-Nu\~{n}ez\textsuperscript{1,$\dagger$}, Maryclare Griffin\textsuperscript{2}, and Lorin Crawford\textsuperscript{3-5,$\dagger$}
\\
\bigskip
\bf{1} Division of Applied Mathematics, Brown University, Providence, RI, USA
\\
\bf{2} Department of Mathematics and Statistics, University of Massachusetts Amherst, Amherst, MA, USA
\\
\bf{3} Microsoft Research New England, Cambridge, MA, USA
\\
\bf{4} Department of Biostatistics, Brown University, Providence, RI, USA
\\
\bf{5} Center for Computational Molecular Biology, Brown University, Providence, RI, USA
\\
\bigskip
$\dagger$ Corresponding E-mails: emily\_winn@brown.edu; lcrawford@microsoft.com 
\end{flushleft}



\section*{Abstract} 

The ability to interpret machine learning models has become increasingly important as their usage  in data science continues to rise. Most current interpretability methods are optimized to work on either (\textit{i}) a global scale, where the goal is to rank features based on their contributions to overall variation in an observed population, or (\textit{ii}) the local level, which aims to detail on how important a feature is to a particular individual in the data set. In this work, a new operator is proposed called the ``GlObal And Local Score'' (GOALS): a simple \textit{post hoc} approach to simultaneously assess local and global feature variable importance in nonlinear models. Motivated by problems in biomedicine, the approach is demonstrated using Gaussian process regression where the task of understanding how genetic markers are associated with disease progression both within individuals and across populations is of high interest. Detailed simulations and real data analyses illustrate the flexible and efficient utility of GOALS over state-of-the-art variable importance strategies.


\section*{Introduction}

Over the past decade, ``interpretability'' has become a  major focus in statistical and probabilistic machine learning. While there remains to be a universal definition for what makes a computational method interpretable \citep[e.g.,][]{guidotti2018survey,carvalho2019machine,hall2019guidelines}, it generally refers to a model's ``ability to explain or to present in understandable terms to a human'' \citep[e.g.,][]{doshi2017towards}. The simple structure of linear models gives an intrinsic interpretation to their parameters and, as a result, enables them to be used for downstream tasks that extend beyond prediction. Part of the utility of linear models is their ability to provide well-calibrated significance measures such as $P$-values, posterior inclusion probabilities (PIPs), or Bayes factors --- all of which lend a notion of statistical evidence about how important each feature is in explaining an outcome variable. Unfortunately, linear models can be underpowered and infeasible to implement in practice. The strict additive assumptions underlying linear regression can be a hinderance in many supervised learning tasks where the variation of a measured response is dominated by nonlinear interactions. As data collection technologies continue to advance, even the most powerful linear models have struggled scale to high dimensions due to both inefficient model fitting procedures \citep[e.g.,][]{Runcie2019fastflexible,Lippert2011FaST,trippe2021for,Lin:2022aa,Schulz:2020aa} and increasingly large combinatorial feature spaces when searching over both additive and non-additive effects \citep[e.g.,][]{crawford_detecting_2017,Stamp:2023aa,pmlr-v97-agrawal19a}. 

Machine learning methods can overcome limitations of linear regression by accommodating nonlinear relationships between features (e.g., through activation units in neural networks or via nonparametric covariance functions in Gaussian processes) and implement scalable training algorithms. However, many machine learning methods are also known to be ``black box'' since they are not inherently transparent about how parameters are learned in making decisions and predicting outcomes \citep[e.g.,][]{rudin2019stop,degrave2021ai,Rudin:2022aa}. Classically, there are two strategies to achieving interpretability of machine learning methods. The first solution attempts to achieve intrinsic interpretability by limiting the architecture of machine learning methods to simple structures \citep{ai2021model}. As an example, in the biomedical sciences, a recent trend has been to develop customized neural network architectures that are inspired by biological systems \citep[e.g.,][]{Demetci:2021aa,elmarakeby2021biologically,Bourgeais2021deepgonet,bourgeais2022graphgonet,Fortelny:2020aa}. Rather having fully connected, potentially over-parameterized architectures, these newer frameworks have partially connected architectures that are based on (\textit{i}) annotations in the literature or (\textit{ii}) derived from relationships between features that have been identified through real-world evidence. In the biomedical application example, each neural network node has an intrinsic interpretation because they encode some biological unit (e.g., signaling pathways, protein motif, or gene regulatory network) and each weight connecting nodes represent known relationships between the corresponding units. A key aspect of this partially connected modeling approach is that it depends on reliable domain knowledge to generate these architectures. When this level of information is not available, as is the case for many practical scientific problems, implementing this strategy can be extremely challenging.

The second strategy to gain interpretability uses \textit{post hoc} or \textit{auxiliary} methods to assess the importance of features after a model has been trained. A wide range of such approaches have been proposed in the literature; however, although many of these techniques share theoretical connections \citep{lundberg2016unexpectedunity}, they generally can be separated into two categories. The first group of methods are ``salience methods'' \citep[also commonly known as ``saliency maps'';][]{simonyan2014deep} which, in their simplest form, provide variable importance by calculating the gradient of a model loss function with respect to each feature for a class of interest. \citet{Kindermans:2019aa} showed that these types of attribution approaches can be highly unreliable in the presence of simple noise structures. In this paper, we will focus on a second class of explainable methods that produce ``sensitivity scores'' which quantify variable importance by measuring the amount predictive accuracy that is lost when a particular feature is perturbed. Common examples in this second class of methods include information criterion \citep{gelman_understanding_2014}, distributional centrality measures \citep{woo2015elucidating,piironen_projection_2016,piironen_comparison_2017,crawford_variable_2019,paananen2019variable,paananen2021uncertainty}, Shapley Additive Explanations (SHAP) \citep{lundberg_unified_2017,chen2022explaining}, and knockoffs \citep{candes_panning_2018,Sesia:2020aa,Sesia:2021aa}. Each of these methods have been shown to have their advantages, but one limitation they all have in common is that they mainly focus on addressing either (\textit{i}) global interpretability where the goal is to rank/select features based on their contributions to overall variation in an observed population, or (\textit{ii}) local interpretability which aims to detail how important a feature is to any particular individual in the data set. In many \red{scientific applications}, it would be ideal to have a measure that leads to conclusions on both scales, simultaneously. For example, \red{in human health}, it is important to understand how a \red{gene is associated with the general progression of a disease} --- but, for the purpose of precision medicine, it is also important to understand how \red{certain genes might have disproportionate effects on individuals coming from different subpopulations \citep[e.g.,][]{martin_clinical_2019,Smith:2022aa}}.

In this work, we present the “GlObal And Local Score” (GOALS) operator: a simple approach that builds off of the distributional centrality literature to provide a measure that assesses both local and global variable importance for features, simultaneously. Our method is entirely general with respect to the modeling approach taken. The only requirements are that we have access to the fitted model and the ability to generate out-of-sample predictions. As a \red{general} illustration of our approach, we focus on using Gaussian process regression. However, also note that this variable importance approach immediately applies to other \red{probabilistic methodologies such as neural networks \citep[e.g., see review in][]{Conard:2023aa}}. We assess our proposed approach in the context of statistical genetics as a way to highlight data science applications that (\textit{i}) contain outcomes that are driven by many covarying and interacting \red{features} \citep[e.g., pairwise interactions between genes;][]{crawford_detecting_2017} and (\textit{ii}) can contain diverse subsets of populations where the importance of features may not be uniform across all individuals in the data. The remainder of the paper is organized as follows. First, we briefly detail the distributional centrality framework for achieving interpretability in nonlinear regression. Here, we review Gaussian processes, motivate the need for an effect size (regression coefficient) analog for features, and define the concept of relative centrality which can be used to perform variable importance. In the next section, we derive the GOALS operator and detail its ability to make local and global interpretations for features. Lastly, we show the utility of our methodology with extensive simulations and a real data analysis of complex traits assayed in a heterogenous stock of mice from Wellcome Trust Centre for Human Genetics \citep{valdar_simulating_2006, valdar_gwas_2006}.


\section*{Overview: Distributional Centrality for Nonlinear Models}

In this work, we will follow positions taken by previous studies and assume that an interpretable statistical method is made up of three key components: (\textit{i}) a motivating probabilistic model, (\textit{ii}) a notion of an effect size (or regression coefficient) for each \red{feature} and (\textit{iii}) a \red{metric that determines the statistical significance of each feature} according to a well-defined null hypothesis \citep{crawford_variable_2019}. The third component is commonly defined by the task of achieving either global or local interpretability. \red{The main objective of global interpretability is to identify features that best explain the variation of an outcome variable within an observed population. In contrast, local interpretability aims to provide an explanation on how important a single \red{feature} is to any particular individual in the data set.} The purpose of this section is to review background which allows us to demonstrat all three of these key components within the context of Bayesian Gaussian process regression for continuous \red{outcomes}; however, note that extending this theoretical framework to other nonlinear methods \citep[e.g., neural networks;][]{Conard:2023aa,ish2019interpreting}, as well as to categorical \red{outcomes} (e.g., binary \red{class labels} in case-control studies), \citep[e.g.,][]{MJ:2011aa} is straightforward. In terms of global interpretability, we will introduce the concept of an effect size analog and describe how distributional centrality measures can be used to perform \textit{post hoc} variable prioritization (also sometimes referred to as performing ``variable importance'' in certain areas of the literature). We then comment on the landscape of existing approaches to assess local interpretability within these same methods and discuss some the need for unifying these concepts for \red{various statistical applications}.

\subsection*{Weight-Space Gaussian Process Regression}

\red{Consider a data set where $\by$ is a continuous $N$-dimensional response vector and $\bX$ is an $N\times J$ design matrix with $N$ observations and $J$ covariates. To build intuition, we begin by specifying a standard linear regression model to analyze the outcome variable such that}
\begin{align}
\by = \bm{f} + \bvarepsilon, \quad \quad \bm{f} = \bX\bbeta, \quad \quad \bvarepsilon \sim\N(\bm{0},\sigma^2\bI),\label{linmodel}
\end{align}
where the function to be estimated $\bm{f}$ is assumed to be a linear combination of \red{the features} in $\bX$ and their respective effects denoted by the $J$-dimensional vector $\bbeta = (\beta_1,\ldots,\beta_J)$ of additive coefficients, $\bvarepsilon$ is a normally distributed error term with mean zero and scaled variance term $\sigma^2$, and $\bI$ denotes an $N\times N$ identity matrix. For convenience, we will assume that the \red{response variable} has been centered and standardized to have mean zero and standard deviation equal to one.

\red{It has been well documented that linear models can be underpowered when the variation of the outcome is driven by non-additive effects \citep[e.g.,][]{Perez-Cruz:2013aa,Yoshikawa:2015aa,cheng_additive_2019}. For example, in genetics applications, nonlinear models have been shown to outperform linear regression in the presence of gene-by-gene interactions \citep{Jiang2019resourceefficient,Weissbrod:2016aa,McCaw:2022aa,Zhou:2022aa}. In these cases, the assumption in Eq.~\eq{linmodel} that the variation in the response $\by$ can be fully explained by additive effects is restrictive. One way to overcome this limitation is to conduct model inference within a high-dimensional function space. In this work, we take a general nonparametric approach and conduct inference in a reproducing kernel Hilbert space (RKHS) by specifying a Gaussian process (GP) prior over the data such that}
\begin{align}
f(\bx)\sim\mathcal{GP}(m(\bx),k_\theta(\bx,\bx^{\prime})),\label{GPprior}
\end{align}
where $f(\bullet)$ is defined by its mean function $m(\bullet)$ (which we will consider to be fixed at zero) and positive definite covariance function $k_\theta(\bullet,\bullet)$. In practice, we assume that our model is only evaluated on the $N$ \red{observations} in our data. When conditioning on these finite \red{samples} (or finite set of locations), the GP prior in Eq.~\eq{GPprior} becomes a multivariate normal distribution \citep{Kolmogorov:1960aa,rasmussen_gaussian_2006} and we can write the following ``weight-space'' nonlinear \red{regression} model 
\begin{align}
\by = \bm{f} + \bvarepsilon, \quad \quad \bm{f} \sim\N(\bm{0},\bK), \quad \quad \bvarepsilon \sim\N(\bm{0},\sigma^2\bI).\label{GPmodel}
\end{align}
Here, $\bm{f} = [f(\bx_1),\ldots,f(\bx_N)]$ is an $N$-dimensional normally distributed random variable with mean vector $\bm{0}$, and the covariance matrix $\bK$ is computed with each element given by $k_{ii^{\prime}} = k_\theta(\bx_i,\bx_{i^\prime})$ where $\bx_i$ and $\bx_{i^\prime}$ denote the \red{features} for the $i$-th and $i^\prime$-th \red{observation}, respectively. Many covariance functions have been shown to implicitly account for higher-order interactions between features, which often lead to more accurate characterization of complex data types \citep{Demetci:2021aa,Wahba:1990aa,tsang2018detecting,tsang2018NIT,crawford_bayesian_2018,Murdoch2019definitions,Cotter:2011aa}. \red{For the demonstrations in the main text of this paper, we will consider $k_{ii^{\prime}}$ to be a nonlinear shift-invariant function.}

Altogether, \red{the ``weight-space'' GP regression model in Eq.~\eq{GPmodel} can be seen as a generalization of the linear model in Eq.~\eq{linmodel} which uses a nonlinear covariance $\bK$ to account for non-additive interactions between features} instead of the usual (additive) gram matrix $\bX\bX^{\T}/J$ \citep[e.g.,][]{Lippert2011FaST,Zhou2012mixedmodel}. \red{Lastly, like linear regression, the GP model can also be easily extended to accommodate fixed effects that are specific to the observations being studied (e.g., age, socioeconomic status) \citep{Campos:2009aa,Shi:2012aa}.} We will not explicitly consider the inclusion of fixed effects here and, instead, will leave those explorations to the reader.

\subsection*{Effect Size Analogs and Relative Centrality Measures}

In this section, we assume access to some trained Bayesian model with the ability to fully characterize or draw samples from its posterior predictive distribution. \red{A central goal in many statistical applications is to jointly infer the true effect size and statistical significance of each feature that is put into the model.} One classic strategy for estimating the regression coefficients in \red{the linear model presented in} Eq.~\eq{linmodel} is to use least squares where the response variable is projected onto the column space of the data $\wh\bbeta := \text{Proj}(\bX,\by) = \bX^{\dagger}\by$ with $\bX^{\dagger}$ denoting some generalized inverse of \red{the design matrix}. We refer to the vector $\wh\bbeta = [\wh\beta_1,\ldots,\wh\beta_J]$ as the (additive) effect size for each \red{feature} in the data set.

The effect size analog was developed with the intention of being the nonparametric version of a regression coefficient for each feature of a nonlinear model \citep{crawford_bayesian_2018}. In general, this leverages the idea that  \red{$\E[\by\cond\bX] = \E[\bm{f}\cond\by]$} when conditioning on $N$ finite observations in Eq.~\eq{GPmodel}. Thus, similar to the linear regression case, the effect size analog can be defined by projecting the smooth nonlinear function onto the column space of the data. While there are many projections one can use \citep[e.g.,][]{woody_model_2021,kowal_fast_2021}, we will consider the following least squares-like projection where
\begin{align}
\wt\bbeta := \text{Proj}(\bX,\bm{f}) = \bX^{\dagger}\bm{f}.\label{ESA}
\end{align}
\red{This is a simple way of understanding the relationships between the features and the response that the nonlinear model has learned.} Under the linear projection, the effect size analogs in Eq.~\eq{ESA} have the usual interpretation. For example, while holding everything else constant, increasing the $j$-th feature by 1 will increase $\bm{f}$ by $\wt\beta_j$ \citep{crawford_bayesian_2018}. Importantly, because of the closed-form projection, drawing samples from the posterior distribution of $\bm{f}$ can be deterministically transformed to samples from the implied posterior distribution of the  effect size analogs.

Similar to regression coefficients in linear models, the effect size analog is not enough on its own to determine variable importance. Indeed, there are many ways to achieve global interpretability based on the magnitude of effect size estimates \citep[e.g.,][]{Barbieri:2004aa,Hoti2006bayesianmapping,Stephens2009bayesianmethods}, but many of these approaches rely on arbitrary thresholding and fail to theoretically test a null hypothesis. One analogy to traditional Bayesian hypothesis testing for nonparametric regression methods is a \textit{post hoc} approach for association mapping via a series of ``distributional centrality measures'' using Kullback–Leibler divergence (KLD) \citep[e.g.,][]{goutis1998model,Smith:2006aa,Tan:2017aa,piironen_comparison_2017,piironen_projection_2016,woo2015elucidating,Alaa:2017aa}. Assume that we have a collection samples from the implied posterior distribution of the effect size analog. We can summarize the importance of the $j$-th \red{feature} in our data by taking the KLD between (\textit{i}) the conditional distribution $p(\wt\bbeta_{-j}\cond\wt\beta_j = 0)$ with the effect of that \red{feature} being set to zero and (\textit{ii}) the marginal distribution $p(\wt\bbeta_{-j})$ with the effect of that \red{feature} having been marginalized over. This is defined by solving the following
\begin{align}
\text{KLD}(j) := \text{KL}\left[p(\wt\bbeta_{-j})\,\|\,p(\wt\bbeta_{-j}\cond \wt\beta_j = 0)\right] = \int_{\wt\bbeta_{-j}}\log\left(\frac{p(\wt\bbeta_{-j})}{p(\wt\bbeta_{-j}\cond \wt\beta_j = 0)}\right) p(\wt\bbeta_{-j}) \diff\wt\bbeta_{-j}.
\label{KLD}
\end{align}
for each $j = 1,\ldots, J$ \red{features} in the data. We can normalize each of these quantities to obtain a final global association metric
\begin{align}
\text{RATE}(j) = \text{KLD}(j)/\sum \text{KLD}(l).\label{RATE}
\end{align}
The above metric is referred to as the ``RelATive cEntrality'' measure or RATE \citep{crawford_variable_2019}. There are two main takeaways that are important about this metric. First, the $\text{KLD}(j)$ value is non-negative, and it equals zero if and only if removing the effect of a given \red{feature} has no impact on explaining the modeled outcome or response (i.e., the posterior distribution of $\wt\bbeta_{-j}$ is independent of $\wt\beta_j$). Second, the RATE measure is bounded on the unit interval [0, 1] with the natural interpretation of providing relative evidence of \red{importance} for each \red{feature} (where values close to 1 suggest greater importance). From a classical hypothesis testing point-of-view, the null under RATE measure assumes that each \red{feature are equally associated with the outcome, while the alternative assumes proposes that some features are much more important than others}. Formally, this can stated as 
\begin{align}
\begin{aligned}
H_0: \text{RATE}(j) = 1/J \quad \text{vs.} \quad H_A: \text{RATE}(j) > 1/J, 
\end{aligned}
\end{align}
where $1/J$ represents the level that all \red{features} in the data have the same relative variable importance. 

\subsection*{Limitations of the Current Distributional Centrality Framework}

There are several notable \red{shortcomings} with effect size analog and RATE framework. First, calculating both the effect size analog and the KLD for each \red{feature} in turn is computationally expensive even with low-rank matrix approximations \citep{crawford_variable_2019}. Both of these operations involve taking inverses of matrices on the order of $J$. As the number of features $J$ grows, these calculations become infeasible. Second, the significance threshold $1/J \rightarrow 0$ as $J\rightarrow\infty$, which effectively means that all \red{variables} will be considered important for high-dimensional settings. Third, while this framework summarizes the global association for each \red{feature} within the observed population, it lacks the ability to locally explain how important \red{variables} are to each \red{individual observation} in the data. This limits its potential impact, \red{for example}, within the context of precision medicine where the goal is to provide individualized patient care. Finally, the least squares projection for the effect size analog in Eq.~\eq{ESA} will only estimate nonlinear effects that are correlated with the linear effects of each \red{feature} \citep{woody_model_2021, kowal_fast_2021,Smith:2023aa}. To see this, define a matrix $\bZ$ whose elements are just each column of $\bX$ squared. Theoretically, we could define quadratic effects by taking the residuals from the regression of $\bm{f}$ on $\bX$ and regressing them onto $\bZ$ in the following way
\begin{align*}
\bgamma = (\bZ^\T\bZ)^{\dagger}\bZ^\T\left(\bI-(\bX^{\T}\bX)^{\dagger}\bX^{\T}\right)\bm{f}.
\end{align*}
Here, implementing the RATE measure on these new effect sizes $\bgamma$ would yield global importance on the quadratic functions of each \red{feature} in the data. However, note that $\bgamma$ vanishes if we combine $\wt\bbeta+\bgamma$ via linear projections onto $\bX$. Therefore, if we wanted to study all linear and quadratic effects together, we would instead need to consider a nonlinear projection such as $\wt\bbeta^2+\bgamma^2$. The projection operator in Eq.~\eq{ESA} will sometimes miss
nonlinear relationships because it only ends up evaluating the part of the nonlinear function $\bm{f}$ that is linearly associated with each \red{feature}. Each of these issues serve as motivation to develop an alternative and more unified framework for nonlinear models.    

\section*{Global and Local Score Operators in Nonlinear Models}

\looseness=-1 We now present a simple alternative to achieve interpretability in nonlinear regression models. We will refer to this new summary as the ``GlObal And Local Score'' (GOALS) operator with the aim to simultaneously identify \red{features} that are significantly associated with \red{a response variable across a} population as well as explain marginal \red{feature} effects on an individual level. Again, let $\bm{f}$ \red{be a function that is estimated from a nonlinear model  (e.g., a Gaussian process)} and consider the scenario where we want to investigate the importance of the $j$-th feature in explaining what that function has learned from the data. To do so, we \red{define perturbed \red{features} $\bX+\bXi^{(j)}$, where $\bXi^{(j)}$ is an $N\times J$ matrix with rows $\bxi^{(j)}$ equal to all zeros except for the $j$-th element which we set to be a vector of some positive constant $\xi$. We then define an $N$-dimensional random variable $\bg^{(j)} = [f(\bx_1 + \bxi^{(j)}),\ldots,f(\bx_N + \bxi^{(j)})]$.} If we think about the interpretation of a regression coefficient in a linear model as detailing the expected change in the mean response given a $\xi$-unit increase in the corresponding covariate (holding all else constant), then a natural quantity to understand the importance of each variable is to study the difference
\begin{align}
\bdelta^{(j)} = \bm{f}-\bg^{(j)}.
\end{align}
Here, each element of the $N$-dimensional vector $\bdelta^{(j)} = (\delta^{(j)}_1,\ldots,\delta^{(j)}_N)$ \red{reflects} the importance of the $j$-th variable for \red{the model} fit with respect to each sample. \red{In other words, it quantifies local variable importance by measuring how much the response changes when a particular feature is perturbed (i.e., similar to the objective of other sensitivity score-based methods).} The sample average $\bar\delta^{(j)} = \sum_i \delta^{(j)}_i/N$ can then be interpreted as a global effect size for the $j$-th variable within the observed population. Intuitively, \red{elements of} $\bdelta^{(j)}$ will be concentrated around zero if the $j$-th \red{covariate} generally has no effect on the \red{response variable that is being analyzed}. This yields the following natural formulation of a null hypothesis for statistical inference and testing
\begin{align}
H_0:  \bdelta^{(j)} = \bm{0} \quad \text{vs.} \quad H_A: \bdelta^{(j)} \ne \bm{0},\label{GOALS}
\end{align} 
where significantly associated variables \red{under the alternative} have \red{$\bdelta^{(j)}$} with magnitudes that largely deviate from zero. Since we are assessing a ``shift'' in function space, each $\bdelta^{(j)}$ takes into account both additive and nonlinear effects for each variable. Note that the GOALS operator can be flexibly implemented by applying the factor $\bXi^{(j)}$ with any constant and even partitioning the data into subsets for which different values of the constant $\xi$ are used. 

\subsection*{\red{Probabilistic Properties of GOALS}}

\red{Although GOALS can be applied to any probabilistic model for the mean  response for arbitrary features, we will demonstrate its properties using a weight-space Gaussian process regression model (e.g., similar to what is detailed in Eq.~\eq{GPmodel})}. To begin, notice that $\bm{f}$ and each $\bg^{(j)}$ are dependent because they are derived from the same set of data $\bX$ and $\by$, respectively. The joint distribution between the $N$-dimensional vectors $\by$, $\bm{f}$, and $\{\bg^{(j)}\}_{j=1}^{J}$ can be specified via the following normal distribution
\begin{align}
\begin{bmatrix}\by \\ \bm{f} \\ \bg^{(1)} \\ \vdots \\ \bg^{(J)}\end{bmatrix} \sim \N\left(\begin{bmatrix} \bm{0} \\ \bm{0} \\ \bm{0} \\ \vdots \\ \bm{0} \end{bmatrix}, \begin{bmatrix} \bA & \bK & \bB^{(1)} & \cdots & \bB^{(J)} \\ 
\bK & \bK & \bB^{(1)} & \cdots & \bB^{(J)} \\ 
\big(\bB^{(1)}\big)^{\T} & \big(\bB^{(1)}\big)^{\T} & \bC^{(1)} & \cdots & \bD^{(1,J)} \\ \vdots & \vdots & \vdots & \ddots & \vdots \\ \big(\bB^{(J)}\big)^{\T} & \big(\bB^{(J)}\big)^{\T} & \bD^{(J,1)} & \cdots & \bC^{(J)} \end{bmatrix} \right),\label{joint_dist}
\end{align}
where $\bA = \bK + \sigma^2\bI$ is the marginal variance of the response vector $\by$; $\bK$ is the variance of $\bm{f}$ using the original \red{design matrix} $\bX$ (as in previous notation); $\bB^{(j)}$ is the covariance between $\bm{f}$ and $\bg^{(j)}$ using the original matrix $\bX$ and the perturbed matrix $\bX+\bXi^{(j)}$; $\bC^{(j)}$ is the variance of $\bg^{(j)}$ using the perturbed matrix $\bX+\bXi^{(j)}$; and $\bD^{(j,l)}$ is the covariance between $\bg^{(j)}$ and $\bg^{(l)}$ having perturbed the $j$-th and $l$-th feature, respectively. We can simplify the above by utilizing the fact that, when the variance function is \red{shift-invariant}, $\bC^{(j)} = \bK$ for all $j$. Using this, we can then derive the following joint distribution between $\bm{f}$ and $\{\bg^{(j)}\}_{j=1}^{J}$, conditioned on the data
\begin{align*}
\resizebox{\hsize}{!}{$\begin{bmatrix}\bm{f} \\ \bg^{(1)} \\ \vdots \\ \bg^{(J)}\end{bmatrix} \bigg|\, \by \sim \N\left(\begin{bmatrix} \bK\bA^{-1}\by \\ \big(\bB^{(1)}\big)^{\T}\bA^{-1}\by \\ \vdots \\ \big(\bB^{(J)}\big)^{\T}\bA^{-1}\by \end{bmatrix}, \begin{bmatrix} \bK - \bK\bA^{-1}\bK & \bB^{(1)} - \bK\bA^{-1}\bB^{(1)} & \cdots & \bB^{(J)} - \bK\bA^{-1}\bB^{(J)} \\ 
\big(\bB^{(1)}\big)^{\T} - \big(\bB^{(1)}\big)^{\T}\bA^{-1}\bK & \bK - \big(\bB^{(1)}\big)^{\T}\bA^{-1}\bB^{(1)} & \cdots & \bD^{(1,J)} - \big(\bB^{(1)}\big)^{\T}\bA^{-1}\bB^{(J)} \\ 
\vdots & \vdots & \ddots & \vdots \\ 
\big(\bB^{(J)}\big)^{\T} - \big(\bB^{(J)}\big)^{\T}\bA^{-1}\bK & \bD^{(J,1)} - \big(\bB^{(J)}\big)^{\T}\bA^{-1}\bB^{(1)} & \cdots & \bK - \big(\bB^{(J)}\big)^{\T}\bA^{-1}\bB^{(J)} \end{bmatrix} \right)$.}
\end{align*}
Lastly, we can write joint distribution for the GOALS operator $\bdelta^{(j)} = \bm{f} - \bg^{(j)}$ in Eq.~\eq{GOALS} as the following
\begin{align}
\begin{bmatrix} \bdelta^{(1)} \\ \vdots \\ \bdelta^{(J)}\end{bmatrix} \bigg|\, \by \sim \N\left(\begin{bmatrix} \left[\bK - \big(\bB^{(1)}\big)^{\T}\right]\bA^{-1}\by \\ \vdots \\ \left[\bK - \big(\bB^{(J)}\big)^{\T}\right]\bA^{-1}\by \end{bmatrix}, \begin{bmatrix} \bSigma^{(1)} & \cdots & \bSigma^{(1,J)} \\ \vdots & \ddots & \vdots \\ \bSigma^{(J,1)} & \cdots & \bSigma^{(J)} \end{bmatrix} \right),\label{GOALS_dist}
\end{align}
where
\begin{align*}
\bSigma^{(j)} &= \bK\bA^{-1}\bK - \big(\bB^{(j)}\big)^{\T}\bA^{-1}\bB^{(j)}-\left[\big(\bB^{(j)}\big)^{\T}-\big(\bB^{(j)}\big)^{\T}\bA^{-1}\bK+\bB^{(j)}-\bK\bA^{-1}\bB^{(j)}\right]\\
\bSigma^{(j,l)} &= \bK - \bK\bA^{-1}\bK + \bD^{(j,l)} - \big(\bB^{(j)}\big)^{\T}\bA^{-1}\bB^{(l)} - \left[\big(\bB^{(j)}\big)^{\T} - \big(\bB^{(j)}\big)^{\T}\bA^{-1}\bK+\bB^{(l)}-\bK\bA^{-1}\bB^{(l)} \right].
\end{align*}
Theoretically, this results in a joint conditional distribution from which to estimate the posterior distribution of each $\bdelta^{(j)}$ and obtain local interpretability. However, in \red{many current data science applications}, where data sets can include hundreds of thousands of \red{observations that have been collected with} millions of \red{features}, it is often desirable to use a more scalable computation than sampling estimates from a full joint distribution. To that end, in this work, we will consider the posterior mean in Eq.~\eq{GOALS_dist} as estimates of local importance and then take the sample means of these values to get a measurement of global importance. More specifically, these two respective values are taken as the following
\begin{align}
\wh\bdelta^{(j)} = \left[\bK - \big(\bB^{(j)}\big)^{\T}\right]\bA^{-1}\by, \quad \quad \bar{\delta}^{(j)} = \sum_i\wh\delta^{(j)}_i/N.\label{sample_mean}
\end{align}
\red{Derivations of the full joint distribution for the global importance scores $[\bar{\delta}^{(1)},\ldots,\bar{\delta}^{(J)}]$, as well as an outline of how to extend GOALS to perform variable importance in probabilistic neural networks, can be found in the Supplementary Material.}

\subsection*{Scalable Computation} 

In practice, we can make use of a few additional matrix algebra properties to efficiently compute estimates from the otherwise computationally intensive distribution outlined in Eqs.~\eq{GOALS_dist} and \eq{sample_mean}. \red{For demonstration, we will assume a Gaussian process with a radial basis covariance function $k_{ii^{\prime}} = \exp\{-\theta\|\bx_i - \bx_{i^\prime}\|^2\}$ where the bandwidth parameter $\theta$ is set using the “median criterion” approach to maintain numerical stability and avoid additional computational costs \citep{Chaudhuri:aa}.} First, it is important to note that the only matrix that needs to be recomputed for each \red{feature} $j$ is the matrix $\bB^{(j)}$ which measures the covariance between the original $\bX$ and the perturbed $\bX+\bXi^{(j)}$. When using the radial basis function, this matrix can be derived for the $j$-th \red{feature} by making the following rank one updates
\begin{align*}
b_{ii^\prime}^{(j)} = k\left(\bx_i,\bx_{i^\prime}+\bxi^{(j)}\right) &= \exp\left\{-\theta\left\|\bx_i-\left(\bx_{i^\prime}+\bxi^{(j)}\right)\right\|^2\right\}\\
&= \exp\left\{-\theta\left[\left\|\bx_i-\bx_{i^\prime}\right\|^2 - 2\left(\bx_i-\bx_{i^\prime}\right)^{\T}\bxi^{(j)} + \left\|\bxi^{(j)}\right\|^2\right]\right\}\\
&= \exp\left\{-\theta\left\|\bx_i-\bx_{i^\prime}\right\|^2\right\}\exp\{-\theta\left[\xi^2- 2\xi\left(x_{ij}-x_{i^\prime j}\right)\right]\}\\
&= k(\bx_i,\bx_{i^\prime})\exp\{-\theta\left[\xi^2- 2\xi\left(x_{ij}-x_{i^\prime j}\right)\right]\},
\end{align*}
where, similar to previous notation, $\bx_i$ and $\bx_{i^\prime}$ are the $i$-th and $i^\prime$-th rows of the \red{design} matrix $\bX$, and $\bxi^{(j)}$ is a row of the matrix $\bXi^{(j)}$ where the $j$-th element is set to some positive constant $\xi$. We can restate the above in matrix notation as
\begin{align}
\bB^{(j)} = \bK \circ \exp\left\{-\theta\left[\xi^2 \bm{1}\bm{1}^{\T} - 2\xi\left(\bx_{\bullet j}\bm{1}^{\T}-\bm{1}\bx_{\bullet j}^{\T}\right)\right]\right\},
\end{align}
where $\bx_{\bullet j}$ is the $j$-th column in the matrix $\bX$ and $\circ$ denotes element-wise multiplication. The main summary is that the computation of each $\bB^{(j)}$ only relies on linear operations after the initial computation of the radial basis covariance matrix $\bK$. These steps extend to other shift-invariant covariance functions (e.g., Laplacian and Cauchy) and a similar rank one update procedure can also be shown for the linear gram matrix (see Supplementary Material). 

\subsection*{Theoretical Connection to Shapley Additive Explanations}

The GOALS operator measures local importance by quantifying the change in function space that occurs when the $j$-th feature of interest is shifted by some nonzero factor. There is a theoretical connection between this strategy and ``SHapley Additive exPlanations'' (SHAP) \citep{lundberg_unified_2017} which is a widely used \textit{post hoc} local interpretability metric in the machine learning literature \citep[e.g.,][]{chen2022explaining}. Briefly, Shapley values assign feature importance weights based on game theoretic principles~\citep{shapley1951notes, roth1988shapley} by essentially determining a payoff for all players when each player might have contributed more or less than the others when attempting to achieve the desired outcome. In applications, this is done by considering all possible subsets of \red{variables} that do not include the $j$-th \red{feature} $\mathcal{S} \subseteq \mathcal{J}\backslash \{j\}$ and then comparing their performance to the performance of a model trained on the same subset as well as the $j$-th \red{feature} $\mathcal{S} \cup \left\{j\right\}$. This weighted average can be represented as the following formula
\begin{align}
\phi_j = \sum_{\mathcal{S}\subseteq \mathcal{J} \backslash \{j\}} \left[\frac{|\mathcal{S}|!(J-|\mathcal{S}|-1)!}{J!}\right]\left(\bm{f}_{\mathcal{S}\cup \{j\}} - \bm{f}_{\mathcal{S}}\right),\label{SHAP}
\end{align}
where $|\mathcal{S}|$ is number of \red{features} in the subset $\mathcal{S}$ and $|\mathcal{J}| = J$ is the total number of \red{features} in the data. Keeping our notation consistent with previous sections, we say that $\bm{f}_{\mathcal{S}\cup \{j\}}$ and $\bm{f}_{\mathcal{S}}$ are the GP regression model fits with and without the $j$-th \red{feature} added to the subset $\mathcal{S}$, respectively.

Rather than removing a given \red{feature} from each subset and calculating model differences, GOALS perturbs each \red{variable} and calculates the corresponding difference in model fit. However, we can relate SHAP to GOALS by considering the special case of a single \red{observation} $N = 1$. In this case, \red{$g^{(j)} = f(\bx+\bxi^{(j)})$} where $\bxi^{(j)}$ is an $1\times J$ vector of all zeros except for the $j$-th element which we set to be some positive constant $\xi$. Note that we can represent the ``shifting'' vector $\bxi^{(j)}$ as the following
\begin{align}
\bxi^{(j)} = \xi\begin{bmatrix}
\mathbbm{1}\{j=1\} & \cdots & \mathbbm{1}\{j=J\},\\
\end{bmatrix}
\end{align}
where $\mathbbm{1}\{\bullet\}$ denotes an indicator function which is returns one for the $j$-th column and 0 otherwise. From this view, we can say that $\mathcal{J}^\prime$ is the set of $J$ indicator random variables which make up elements of $\bxi^{(j)}$. We can also therefore rewrite the GOALS operator as 
\begin{align}
\delta^{(j)} = f - g^{(j)} = f_{\mathcal{J}} - f_{\mathcal{J}\cup\mathcal{J}^{\prime}}.
\end{align}
If we set $\xi = -x_{j}$, the GOALS operator behaves similarly to a SHAP value, as $\bg^{(j)}$ represents the model fit where the $j$-th covariate is set to zero. In this case, GOALS could be seen as an approximation to SHAP where GOALS only considers the single subset of $J-1$ \red{features}, excluding the $j$-th \red{feature}, whereas SHAP considers all every possible subsets of \red{variables} that do not include the $j$-th \red{feature (which can be computationally intensive for large data sets)}. 

Lastly, it is worth noting that there are scenarios where we would expect GOALS and SHAP to provide different local interpretability rankings for the $j$-th covariate. The factorial in the SHAP weight computation in Eq.~\eq{SHAP} favors both the smallest and largest subsets of $\mathcal{J}$ and penalizes subsets $\mathcal{S}$ of the size $|\mathcal{S}|\approx J/2$. This means that if the $j$-th \red{feature} has an effect on the \red{response} via marginal effects, then both GOALS and SHAP are likely to give that \red{feature} a high ranking. If the $j$-th \red{feature} is only influential on a \red{response} through a moderate number of interactions (i.e., within sets of size $J/2$), then GOALS may rank that \red{feature} higher (relative to other \red{features}) than SHAP will. However, on the other hand, if the $j$-th \red{feature} is influential through pairwise interactions with nearly all other \red{features in the data set}, then SHAP may provide a higher relative rank for that \red{feature} than GOALS \red{--- although, this scenario is probably least likely to happen in practice}. Furthermore, SHAP may rank \red{variables} that are highly correlated with each other lower than GOALS --- this is because the difference in model fits $\bm{f}_{\mathcal{S}\cup \{j\}} - \bm{f}_{\mathcal{S}}$ may be small when \red{feature} $j$ is highly correlated with \red{features} in $\mathcal{S}$. We show that these expectations are supported empirically in the next section.


\section*{Results}

We now illustrate the benefits of our simple approach for global and local interpretability in extensive simulations and real data analyses. First, we conduct a proof-of-concept simulation study to help the reader build a stronger intuition for how GOALS prioritizes influential variables on both a local and global scale, simultaneously. To provide concrete points of reference, we will also show how the Shapley Additive Explanations (SHAP) \citep{lundberg_unified_2017} approach assigns feature importance weights locally and we will demonstrate how the distributional centrality framework using the effect size analog with RATE performs global interpretability \citep{crawford_bayesian_2018,crawford_variable_2019}. We also show that GOALS is much more scalable than both methods as both the number of observations and genetic markers increase. For the second analysis in this section, we implement a more realistic simulation scheme to assess how GOALS performs association mapping compared to various \textit{post hoc} variable importance, Bayesian shrinkage, and regularization modeling techniques. Lastly, we apply the GOALS operator to six quantitative traits assayed in a heterogenous stock of mice from Wellcome Trust Centre for Human Genetics \citep{valdar_simulating_2006, valdar_gwas_2006}.

\subsection*{Simulation Studies}

The general design of the following simulation studies \red{has been previously used} to explore the power of \red{variable importance methods \citep{crawford_bayesian_2018,crawford_variable_2019,Demetci:2021aa,Smith:2023aa}. Once again, let $\bX$ be a design matrix of $N$ observations with $J$ \red{features}. To generate synthetic data, we select a subset of causal features from the design matrix and then use the following linear model
\begin{align}
\by = \sum_{c\in \mathcal{C}}\bx_c\beta_c+\bW\btau+\bZ\bomega+\bvarepsilon, \quad \quad \bvarepsilon\sim\N(\bm{0},\sigma^2\bI).\label{sim_model}
\end{align}
where $\by$ is an $N$-dimensional synthetic response vector; $\mathcal{C}$ represents the set of all randomly selected causal features; $\bx_c$ is the $c$-th causal feature vector with a corresponding nonzero additive effect size $\beta_c$; $\bW$ is an $N\times M$ dimensional matrix which holds all pairwise interactions between the causal features, with the columns of this matrix assumed to be the Hadamard (element-wise) product between feature vectors of the form $\bx_j\circ\bx_k$ for the $j$-th and $k$-th features; $\btau$ is the $M$-dimensional vector of interaction effect sizes; $\bZ$ contains covariates representing additional population structure between the samples in the data with corresponding effects $\bomega$; and $\bvarepsilon$ is an $N$-dimensional vector of environmental noise. For simplicity, we will consider $\bZ$ to be the top ten principal components (PCs) from the design matrix $\bX$. In these simulations, we assume that the total variation of the synthetic response variable is $\V[\by] = 1$. We allow the additive and interaction effect sizes to be randomly drawn from standard normal distributions. Next, we scale the additive, pairwise interactions, population structure, and the environmental noise terms so that they collectively explain a fixed proportion of the total variance where
\begin{align}
\V\left[\sum_{c\in \mathcal{C}}\bx_c\beta_c\right] = \rho v^2, \quad \quad \V[\bW\btau] = (1-\rho)v^2, \quad \quad \V[\bZ\bomega]+\V[\bvarepsilon] = 1-v^2. 
\end{align}
Intuitively, $v^2$ determines how much variance in the simulated response is due to signal versus noise, while $\rho$ is a mixture parameter which determines how much of the signal is driven by additive versus interaction effects. Below, we will consider studies where $v^2 \in \{0.3, 0.6\}$. We will also assess different cases by setting $\rho \in \{0.5, 1\}$, where the former assumes that additive and interaction effects contribute equally to the total variation in the response, and the latter assumes only additive effects contribute to the signal.}

\paragraph{Proof-of-Concept Simulations: Low-Dimensional Analysis.} In this subsection, we provide a low-dimensional proof-of-concept simulation study \red{(i.e., $N > J$)}. To accomplish this, we generate \red{synthetic data $\bX$ with $N =$ 2000 observations and $J = 25$ covariates where each feature is drawn from a standard normal distribution}. In these simulations, we generate synthetic \red{outcome variables} using Eq.~\eq{sim_model} by fixing the \red{signal-to-noise ratio to be $v^2 = 0.6$ and omitting population structure effects by setting $\bomega = \bm{0}$.} Here, we assume some subset of the \red{features} $\mathcal{C} = \{8,9,10,23,24,25\}$ to be causal. We then consider five different simulation scenarios:
\begin{itemize}
\item \textbf{Scenario I (Additive and Interaction Effects):} The subset $\{23,24,25\}\subseteq\mathcal{C}$ are causal \red{features}, where all three have additive effects and \red{features} \#23 and \#24 interact with \#25, respectively. 
\item \textbf{Scenario II (Additive and Interactions Effects from Different \red{Groups}):} All \red{features} in the set $\mathcal{C}$ are causal. \red{Features} \#8-10 only have interaction effects and \red{features} \#23-25 only have additive effects. Specifically, \red{features} \#8 and \#9 each interact with \#10, separately. 
\item \textbf{Scenario III (Overlapping Additive and Interaction Effects):} All \red{features} in the set $\mathcal{C}$ are causal. \red{Features} \#23-25 each have additive effects; while, \red{feature} \#8 interacts with \#10 and \#9 interacts with \#25, respectively.
\item \textbf{Scenario IV (Interaction Effects Only):} All \red{features} in the set $\mathcal{C}$ are causal only through interaction effects. \red{Features} \#8 and \#9 each interact with \#10, separately; while, \red{features} \#23 and \#24 each interact with \#25, separately.
\item \textbf{Scenario V (Noise Only):} None of the \red{features} in the data have an association with the \red{response}. Represents the case when assumptions of the null model are met. 
\end{itemize}
We want to point out that, while this is indeed a small proof-of-concept study, each of these cases highlight settings that we might experience \red{in} real \red{applications}. For each scenario, we fit a standard GP regression model similar to Eq.~\eq{GPmodel} under a zero mean prior and a radial basis covariance function. 

\begin{figure}[H]
\centering
\includegraphics[width = \textwidth]{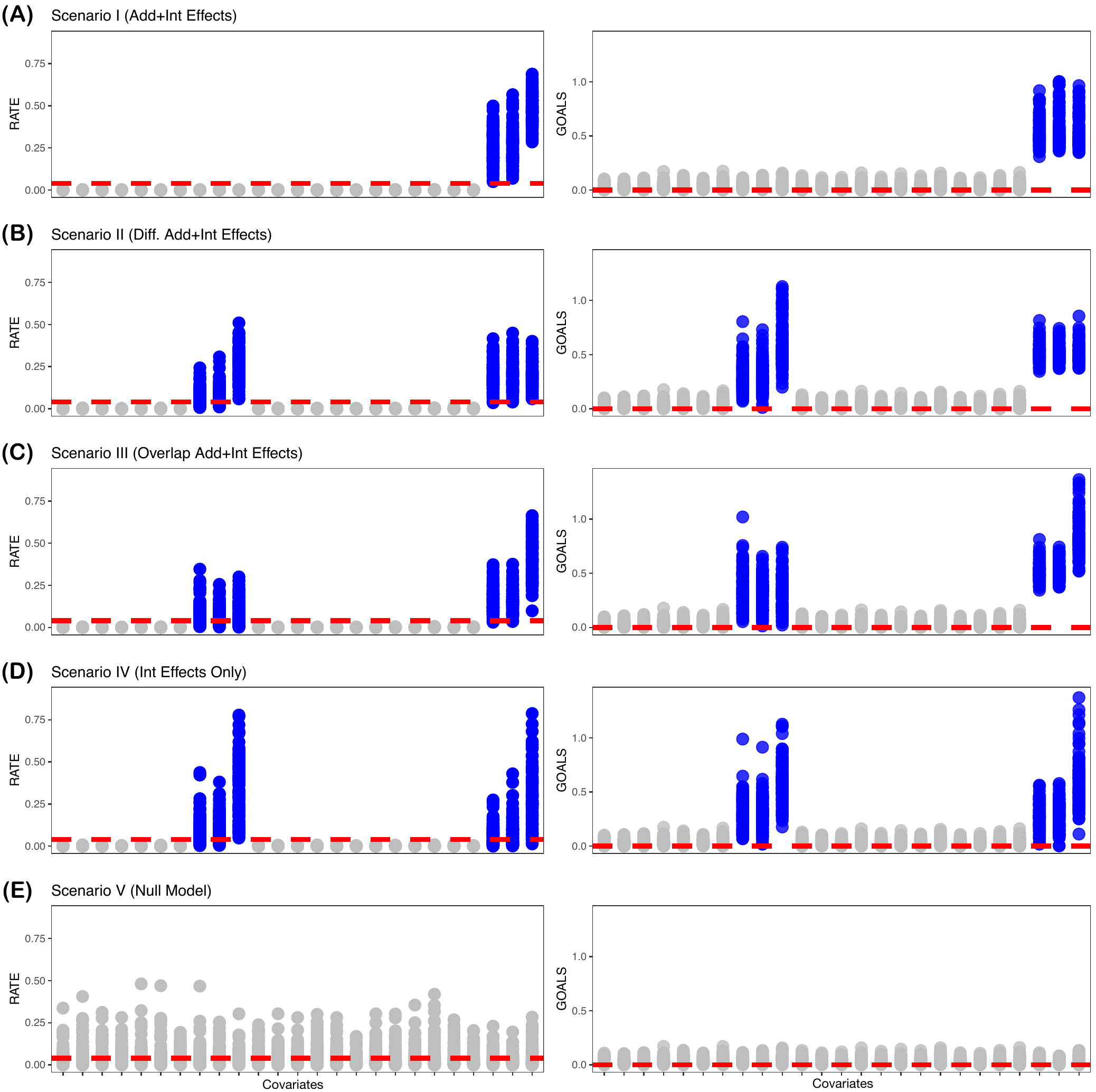}
\caption{\textbf{Proof-of-concept simulations to demonstrate how GOALS and RATE globally prioritize \red{important variables} with varying degrees of additive and interaction effects.} These simple simulations assume that synthetic \red{responses have a signal-to-noise ratio equal to $v^2 = 0.6$ with $(1-\rho)$ = 0\% to 50\% of the signal stemming from interaction effects. Points highlighted in blue are covariates that have nonzero effects within each of the five different scenarios.} To compare the null hypotheses for the two approaches, we also display red dashed lines that are drawn at the level of relative equivalence (i.e., $1/J$) for RATE (left column) and at zero for GOALS (right column), respectively. Note that the scales of the y-axes are different because RATE is theoretically bounded on the unit interval [0, 1]. Here, the main takeaway is that, because the GOALS operator measures variable importance in function space, it is more robust to identifying \red{features} whose associations are driven primarily by interaction effects. All results shown in this figure are based on 100 replicates.}
\label{Fig1}
\end{figure}

Figure \ref{Fig1} contains the global variable importance results for GOALS \red{using perturbation parameter $\xi = 1$} and RATE on Scenarios I-V for 100 simulated replicates. Here, we perform RATE on a GP model using effect size analogs computed with the linear projection as in Eqs.~\eq{ESA}-\eq{RATE}, while the GOALS operator is calculated on the GP model as in Eq.~\eq{sample_mean}. In Figure \ref{Fig1}, the known causal \red{features} for each scenario are colored in blue. To compare the null hypotheses for the two approaches, we also display red dashed lines that are drawn at the level of relative equivalence (i.e., $1/J$) for RATE and at zero for GOALS, respectively. \red{For the alternative simulation Scenarios I-IV, any causal variables with importance scores above the significance thresholds $1/J$ and 0 are considered to be true positives for RATE and GOALS --- all other variables with importance scores above these thresholds are false discoveries. In the null simulation Scenario V, all variables should appear below the respective significance thresholds for both methods.} Overall, we see that both methods perform similarly in identifying causal \red{features} that have both additive and interaction effects on the \red{response} (Figure \ref{Fig1}A). However, GOALS proves to be a better discriminator between causal and non-causal \red{features} than RATE when interaction effects occur in isolation (i.e., \red{covariates} are involved in an interactions without necessarily having an additive effect). Importantly, GOALS exhibits a more robust control of the false negative rate in exchange for a slight increase in false discovery for these scenarios (see how the RATE and GOALS operators relate to the null threshold lines in Figures \ref{Fig1}B-D). This result highlights the potential limitation of the linear projection that RATE uses to compute the effect size analog and demonstrates its potential to miss associations that stem from nonlinear interactions (especially when \red{features} only have non-additive effects such as \red{variables} \#8 and \#9). Lastly, GOALS is better calibrated when \red{data} are generated from complete noise (i.e., when there are no true associations between \red{features  $\bX$ and outcome $\by$}) (Figure \ref{Fig1}E). This is due to the fact that the GOALS operator assesses the global importance variables based on their individual contribution to the model fit. While the concept of relative centrality is intuitive, achieving a completely uniform distribution of RATE values at $1/J$ under the null model will rarely happen in practice (especially in applications where spurious associations between correlated features and the modeled response can occur). In other words, due to the stochastic nature of data, one variable will always appear relatively more important than another which can lead to ill-informed analyses during downstream tasks under the RATE framework.

\begin{figure}[ht]
\centering
\includegraphics[width = 0.84\textwidth]{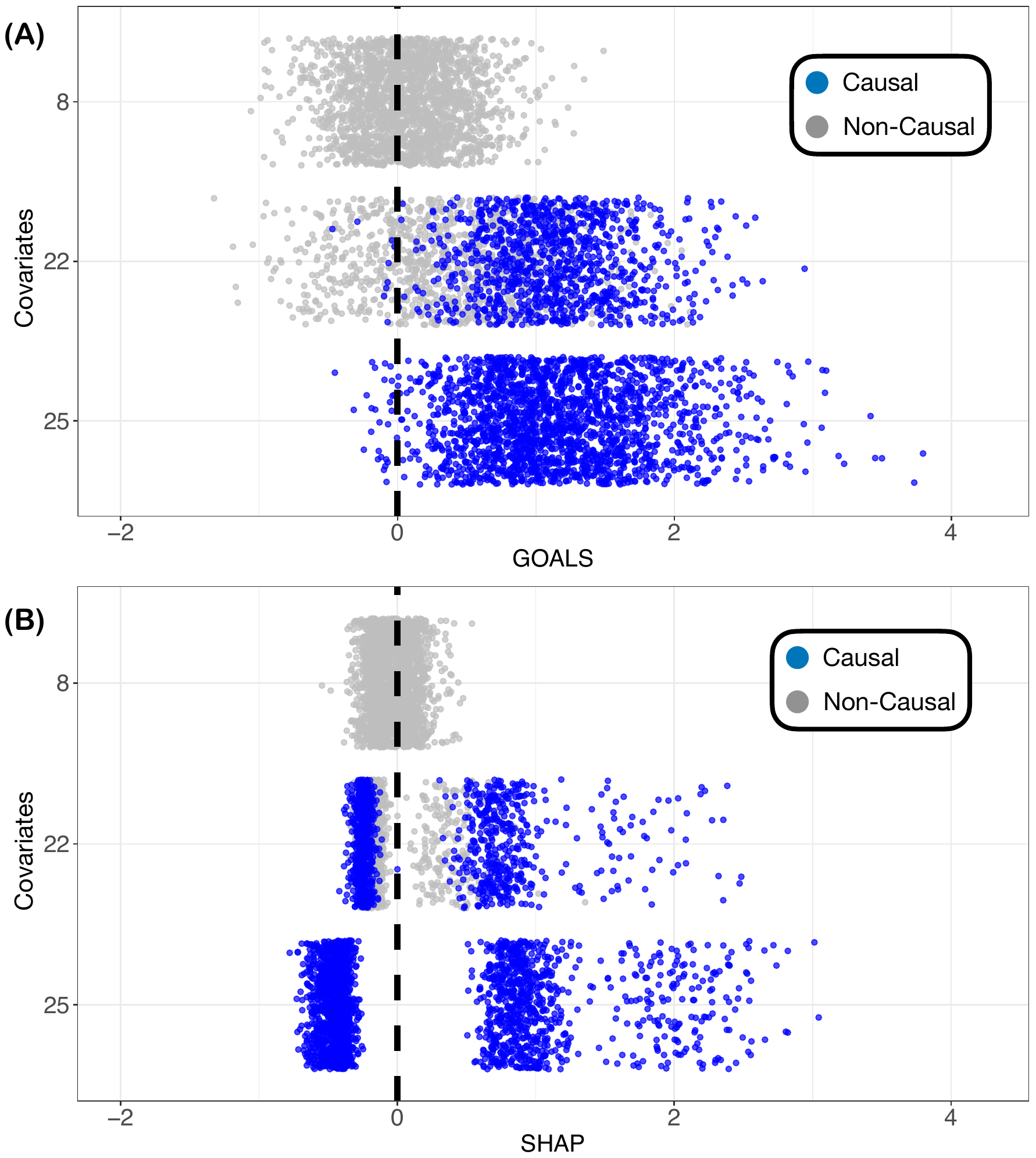}
\caption{\textbf{Proof-of-concept simulations to demonstrate how GOALS and SHAP locally prioritize \red{important variables} that have varying level of effects on specific subsets of the population.} These simple simulations assume that synthetic \red{responses have a signal-to-noise ratio equal to $v^2 = 0.6$ with $(1-\rho)$ = 0\% to 50\% of the signal stemming from interaction effects. Points highlighted in blue are covariates that have nonzero effects within each of the five different scenarios.} Here, each point is an individual. We highlight the local variable importance metrics for three specific \red{features} according \textbf{(A)} GOALS and \textbf{(B)} Shapley Additive Explanations (SHAP). In this simulation study, \red{covariate} \#8 is null feature and does not contribute to the phenotypic variation; \red{covariate} \#25 has additive and interaction effects that are associated with all individuals; and \red{covariate} \#22 has an additive effect for only half of the population. A point is blue if the corresponding labeled \red{covariate} has a nonzero effect for that individual. The main takeaway of this analysis is that, in the case where a \red{feature} has an effect on \red{the response} for only a subset of the observed population, the local distribution of the SHAP and GOALS operators will be multimodal allowing for individualized summaries of \red{variable importance} on specific observations. The black dashed line is drawn at zero to represent a threshold where a \red{feature} has no effect for a given sample.}
\label{Fig2}
\end{figure}

Another major contribution of GOALS is that it also provides local explanations of how variables affect model fit for each individual in the data. For example, in \red{biomedical} applications, this can yield key insight in the event that a \red{gene} is biomarker for only a specific subset a population. To demonstrate the utility of GOALS in this case, we consider a sixth simulation scenario where
\begin{itemize}
\item \textbf{Scenario VI (Population Specific Effects):} The subset $\{22, 23,24,25\}\subseteq\mathcal{C}$ are causal \red{features}. \red{Features} \#23-25 have additive and interaction effects that are associated with all individuals; while, \red{feature} \#22 has an additive effect for only half of the population.
\end{itemize}
Figure \ref{Fig2} shows the distribution of the local individual-level GOALS operator for \red{variables} \#8, \#22, and \#25 in this split scenario. As a baseline, we also show results from running a local analysis with SHAP. For clarity, \red{variable} \#8 is a non-causal \red{feature} in this scenario. There are a few key takeaways in this empirical illustration. First, when a \red{feature} has a no effect on the \red{response}, the distribution of the local scores for both GOALS and SHAP are centered at 0. Conversely, \red{features} with nonzero effects on the \red{outcome} have GOALS and SHAP operators with magnitudes that are centered distinctly away from the origin. One difference here is that the GOALS values tend to have the same sign, while the SHAP metric can be positive or negative. In the case where a \red{feature} has an effect for only a subset of the observed population, the local distribution of the SHAP and GOALS operators will be multimodal allowing for individualized summaries of \red{variable importance} on specific observations. In Figure \ref{Fig2}, this characteristic is more distinct with the GOALS operator where there is clearer separation in values for \red{variable} \#22 in \red{samples} where it has a nonzero effect. 

\paragraph{Method Comparisons: High-Dimensional Global Variable Importance.} We now assess the power of GOALS and its ability to effectively prioritize causal variables in high-dimensional data settings. In this analysis, we use \red{real data as our design matrix $\bX$ to generate a synthetic outcome $\by$. Here, we take} data from the Wellcome Trust Case Control Consortium (WTCCC) 1 study which initially consisted of 2,938 \red{samples} with 458,868 \red{genetic features. The features in this data are known as single nucleotide polymorphisms (SNPs), each of which are originally encoded as {0, 1, 2} copies of a reference allele at each locus. We follow the same quality control procedures used in previous studies \citep{WTCCC}.} Missing \red{data} were imputed by using the BIMBAM software \citep[\url{http://www.haplotype.org/bimbam.html};][]{Servin:2007aa}. In these simulations, we use all \red{features} with minor allele frequencies (MAFs) above 1\% on chromosome 22 to generate continuous \red{outcome variables}. \red{After preprocessing, all genetic features were centered and scaled to have mean zero and standard deviation equal to one.} Exclusively considering this group of individuals and SNPs resulted in a final data set consisting of $N =$ 2,938 samples and $J =$ 5,747 \red{features}.

During each simulation run, \red{we randomly choose a set of $|\mathcal{C}| = 30$ causal SNPs.} Next, we set the \red{signal-to-noise ratio $v^2 =0.3$} and consider two choices for the \red{contribution stemming from interactions between causal features} $\rho \in \{0.5, 1\}$. We also consider simulations with and without population \red{structure} effects by allowing the top ten principal components (PCs) \red{from the design matrix $\bX$ to} make up to 10\% of the overall variation in the synthetic \red{outcome variable $\by$}. In total, this resulted in four scenarios based on different parameter combinations: (\textit{i}) $\rho =$ 1 and $\V[\bZ\bomega]$ = 0; (\textit{ii}) $\rho =$ 1 and $\V[\bZ\bomega]$ = 0.1; (\textit{iii}) $\rho =$ 0.5 and $\V[\bZ\bomega]$ = 0; and (\textit{iv}) $\rho =$ 0.5 and $\V[\bZ\bomega]$ = 0.1. In other words, scenarios I and II consider \red{data} with only additive effects; while, scenarios III and IV consider \red{data} with both additive and interaction effects. Additionally, scenarios II and IV have the additional complexity of having nonzero \red{effects from} population \red{structure} which is not observed in scenarios I and III.  

We compare the global power of the GOALS measure to a list of \red{variable importance} techniques. Specifically, these methods include: (a) the \textit{post hoc} framework of estimating effect size analogs for the features used in a GP regression model and determining their importance using distributional centrality via RATE \citep{crawford_variable_2019}; (b) a univariate linear model (SCANONE) \citep{Yandell:2007aa}; (c) L1-regularized ``least absolute shrinkage and selection operator'' (LASSO) regression \citep{tibshirani1996lasso}; (d) the combined regularization utilized by the Elastic Net \citep{zou2005regularization}; \red{(e) a random forest (RF) \citep{Ishwaran:2019aa} fit with 500 trees; a gradient boosting machine (GBM) \citep{Friedman:2001aa} fit with 100 trees; and a Bayesian additive regression tree (BART) \citep{Chipman:2010aa} fit with 200 trees and 1000 Markov chain Monte Carlo (MCMC) iterations.} Note that SCANONE produces $P$-values, and the LASSO and the Elastic Net give magnitudes of regression coefficients. The latter two regularization approaches were fit by first learning tuning parameter values via 10-fold cross validation. \red{Additionally, features in the RF and GBM are ranked by assessing relative influence which is computed by taking the average total decrease in the residual sum of squares after tree splitting on each variable; while, in BART, features are ranked by the average number of times that they are used in decisions for each tree.} Indeed, the SHAP value framework can also be used for \textit{post hoc} assessment of global interpretability by taking the average of local scores across observations for each feature in the data. However, because the SHAP approach considers all possible subsets of \red{features} when determining variable importance, it does not scale well to high-dimensional settings. For this reason, we do not consider it for comparison in this simulation study \red{(see next section for its application to a subset of real data)}.

\begin{figure}[ht]
\centering
\includegraphics[width = \textwidth]{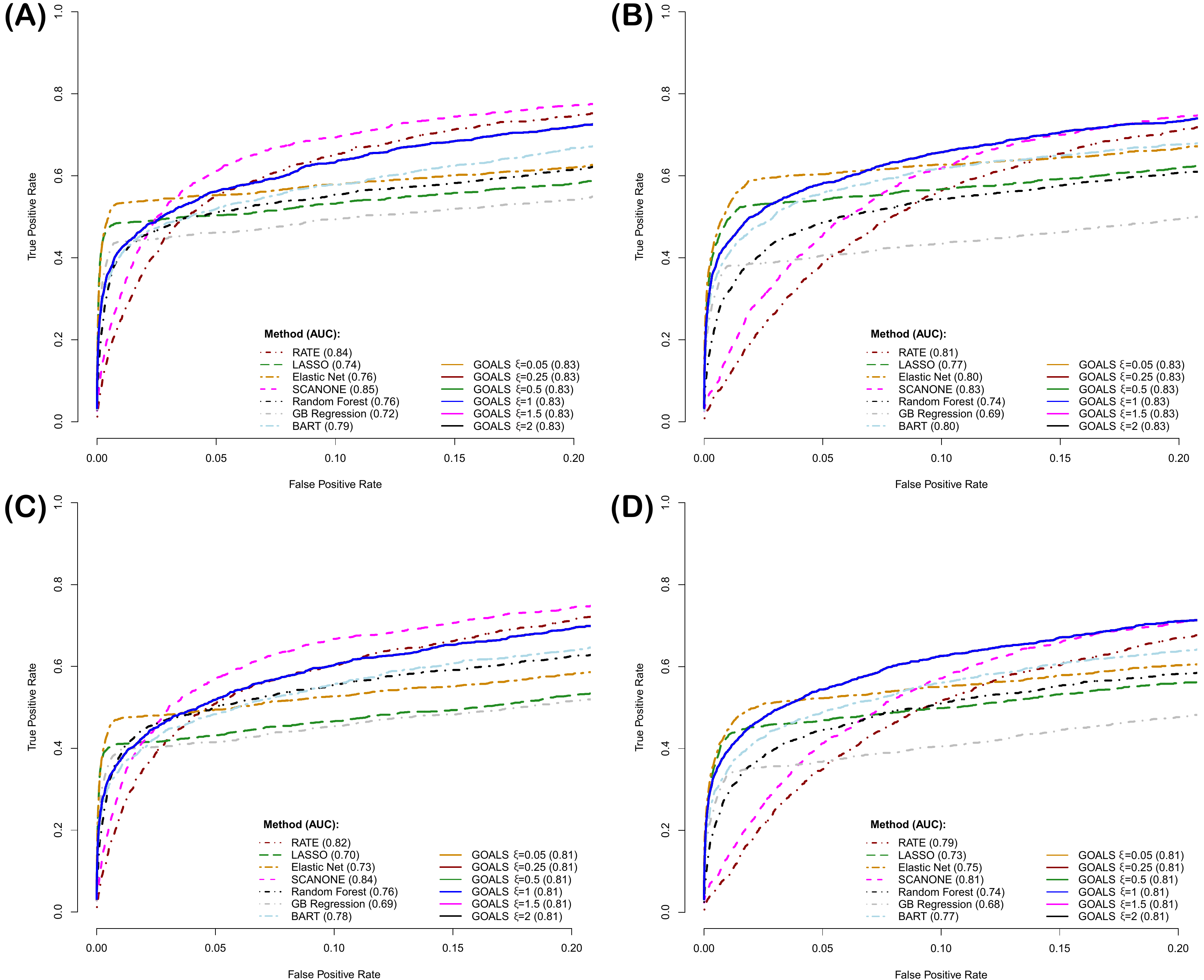}
\caption{\textbf{Receiving operating characteristic (ROC) curves comparing the performance of GOALS against other global variable importance approaches in simulations.} Here, synthetic \red{responses are simulated to have a signal-to-noise ratio equal to $v^2 = 0.6$} with only additive effects in panels \textbf{(A)} and \textbf{(B)}, and a combination of additive and pairwise interaction effects in panels \textbf{(C)} and \textbf{(D)}. This is controlled by a free parameter $\rho = \{0.5, 1\}$ which was used to determine the proportion of \red{signal} that is contributed by additivity. The \red{response variables} simulated in panels \textbf{(B)} and \textbf{(D)} also have the additional complexity of having population stratification effects. We show results using \red{Gaussian process regression with GOALS across a wide range of values for the perturbation parameter $\xi$.} Competing approaches include: Gaussian process regression with RATE (red), LASSO regularization (green), the Elastic Net (yellow), the SCANONE method (pink), \red{a random forest (RF) (black), a gradient boosting machine (GBM) (grey), and a Bayesian additive regression tree (BART) (light blue). Methods using GOALS are illustrated as a solid lines, while the competing models are shown as dotted lines. Note that the performance of GOALS is not sensitive to the choice of $\xi$ in these simulations, so all solid lines fall on top of each other. Lastly, note} that the upper limit of the x-axis (i.e., false positive rate) has been truncated at 0.20. All results are based on 100 simulated replicates.}
\label{Fig3}
\end{figure}

Each method is evaluated based on its ability to effectively prioritize causal features in 100 different simulated data sets. \red{We consider the 30 variables in the causal set $\mathcal{C}$ to be true positives and all other variables to be true negatives. The criteria we use compares the false positive rate (FPR) with the rate at which true causal variables are selected first by each model (TPR)}. \red{This information is depicted as receiver operating characteristic (ROC) curves in Figure \ref{Fig3}. Specifically, for each method, we rank features from most to least important. Starting with the top ranked variable, we then use a sliding threshold to create a set of ``selected'' features. During each iteration, we compute the number of true and false positives in the selected set (which we will denote as TP and FP, respectively). We then calculate the TPR and FPR as the following
\begin{align}
\text{TPR} = \text{TP}/|\mathcal{C}|, \quad \quad \text{FPR} = \text{FP}/(J-|\mathcal{C}|),
\end{align}
where, again, $|\mathcal{C}| = 30$ is the number of causal variables in the simulation and $J-|\mathcal{C}|$ then represents the number of true negatives. Figure \ref{Fig3} illustrates the mean ROC curve across all 100 replicates per simulation scenario,} where the upper limit of the FPR on the x-axis has been truncated at 0.2. This is further quantified by assessing the entire area under the curve (AUC) in the legend for further comparison --- \red{where a higher AUC points to better model performance}. 

Overall method performance varies depending on the two factors: (a) the presence of interaction effects, and (b) additional structure due to population stratification. For example, most methods perform best in the first simulation scenario where \red{data is generated by causal variables with} only additive effects (e.g., Figure \ref{Fig3}A). This power generally decreases in the presence of \red{population structure} (e.g., Figure \ref{Fig3}B) or when \red{causal variables are involved in pairwise interactions} (e.g., Figure \ref{Fig3}C and \ref{Fig3}D). GOALS outperforms LASSO, Elastic Net, \red{RF, GBM, and BART} consistently in every scenario and performs competitively with SCANONE and RATE in every scenario. More specifically, GOALS is a top performer in scenarios with additional population \red{structure} at lower false positive rates. While RATE performs generally well in each of these scenarios, the algorithm often takes much longer than GOALS to run as the number features increases. For a data set with $J$ = 500 features and $N$ = 1000 samples, RATE has an average runtime of 60 seconds on computing cluster with 30 nodes whereas GOALS takes only a second to complete. \red{Lastly, to illustrate the robustness of GOALS, we apply our method while using a range of values for $\xi = \{0.05, 0.25, 0.5, 1, 1.5, 2\}$ and show that the performance of GOALS is relatively robust to the choice made for this parameter.} We argue that these simulations highlight GOALS as a reliable option for interpretability given its consistent performance across a wide range of scenarios and its scalability as data sizes increase. GOALS also has the additional benefit of allowing for local variable importance analyses which is something that RATE and SCANONE do not provide.

\subsection*{Global and Local Association Mapping in Heterogenous Stock of Mice}

In this section, we apply GOALS to genetic data from a heterogenous stock of mice collected by the Wellcome Trust Centre of Human Genetics (\url{http://mtweb.cs.ucl.ac.uk/mus/www/mouse/index.shtml}) \citep{valdar_simulating_2006, valdar_gwas_2006}. The genotypes from this study were downloaded directly using the \texttt{BGLR-R} package \citep{perez_bglr_2014}. This study contains $N =$ 1,814 heterogenous stock of mice from 85 families (all descending from eight inbred progenitor strains) and 131 quantitative traits that are classified into 6 broad categories including behavior, diabetes, asthma, immunology, haematology, and biochemistry. Phenotypic measurements for these mice can be found freely available online to download (details can be found at \url{http://mtweb.cs.ucl.ac.uk/mus/www/mouse/HS/index.shtml}). In this study, we focus on three of these complex traits: body weight, percentage of CD8+ cells, and high-density lipoprotein (HDL) content. Each of these phenotypes were previously corrected for sex, age, body weight, season, and year \citep{valdar_simulating_2006, valdar_gwas_2006}. For individuals with missing genotypes, we imputed values by the mean genotype of that SNP in their corresponding family. Only polymorphic SNPs with minor allele frequency above 5\% were kept for the analyses. This left a total of $J =$ 10,227 \red{genetic features} that were available for all mice. 

We chose to analyze this particular data set for a few reasons. The first reason is that RATE has been previously applied to these same three traits to perform nonlinear \textit{post hoc} variable importance \citep{crawford_variable_2019} --- thus, it provides a methodological baseline for the performance of GOALS on the global level. The second reason is that common environmental effects caused by the mice sharing the same cage have been shown to have nonzero contribution to the overall variance observed in these traits \citep{crawford_bayesian_2018}. Therefore, it means that one might expect to observe varying local SNP effects between mice assigned to different cages. Lastly, the mice in this study are known to be genetically related and the measured have varying levels of broad-sense heritability \red{(i.e., signal-to-noise ratios)} with nonzero contributions from both additive and non-additive genetic effects \citep{valdar_gwas_2006, chen_number-of-X_2012}. As result, this data set represents a realistic mixture of the simulation scenarios we detailed in the previous sections. 

\begin{landscape}

\begin{figure}[ht]
\centering
\vspace*{-0.7 in}
\hspace*{-0.3 in}
\includegraphics[scale = 0.65]{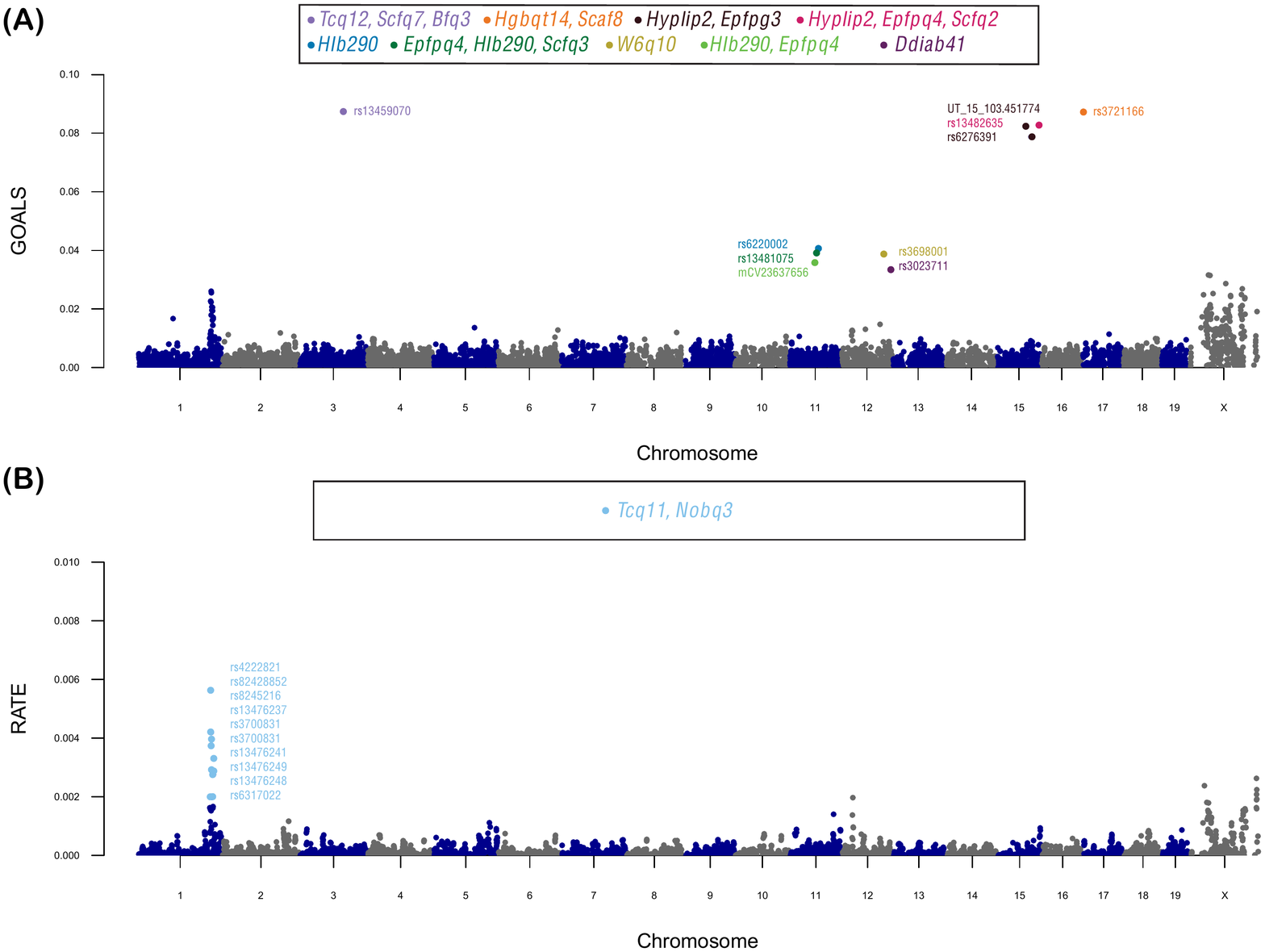}
\caption{\textbf{Manhattan plot of variant-level association mapping results for high-density lipoprotein (HDL) content in the heterogenous stock of mice data set from the Wellcome Trust Centre of Human Genetics \citep{valdar_simulating_2006, valdar_gwas_2006}.} Panel \textbf{(A)} depicts the global GOALS measure \red{(with $\xi = 1$)} of quality-control-positive SNPs plotted against their genomic positions after running a Bayesian Gaussian process (GP) regression on the quantitative trait. As a direct comparison, in panel \textbf{(B)}, we also include results after implementing RATE on the same fitted GP model. In this figure, chromosomes are shown in alternating colors for clarity. The top 10 highest ranked SNPs by GOALS and RATE, respectively, are labeled and color coded based on their nearest mapped gene(s) as cited by the Mouse Genome Informatics database (\url{http://www.informatics.jax.org/}) \citep{bult_mouse_2019}. These annotated genes are listed in the legends of each panel. \red{A comparison of these results to a random forest (RF), a gradient boosting machine (GBM), and a Bayesian additive regression tree (BART) can be found in Figure \ref{Fig_S1}. A complete list of the variable importance values provided by each method} for all SNPs can be found in Table \ref{Tab_S1}.}
\label{Fig4}
\end{figure}

\end{landscape}

For each trait, we fit a GP regression model with GOALS and RATE, \red{a random forest (RF) \citep{Ishwaran:2019aa} fit with 500 trees, a gradient boosting machine (GBM) \citep{Friedman:2001aa} fit with 100 trees; and a Bayesian additive regression tree (BART) \citep{Chipman:2010aa} fit with 200 trees and 1000 Markov chain Monte Carlo (MCMC) iterations}. Figures \ref{Fig4} and \red{\ref{Fig_S1}-\ref{Fig_S5}} display the variant-level mapping results \red{via Manhattan plots} after assessing variable importance using GOALS \red{with $\xi = 1$}, RATE, RF, GBM, and BART in HDL content, body weight, and the percentage of CD8+ cells, respectively. \red{In these plots, larger values mean ``enrichment'' in a given genomic position.} Notable SNPs are annotated and color coded according to their nearest mapped gene(s) as cited by the Mouse Genome Informatics database (\url{http://www.informatics.jax.org/}) \citep{bult_mouse_2019}. We also provide summary tables which lists the corresponding \red{variable importance scores} for all SNPs (see Tables \ref{Tab_S1}-\ref{Tab_S3} in the Supplementary Material). In general, GOALS demonstrated the ability to identify \red{biologically relevant} \red{signal in all three traits that were missed by the other competing approaches. This was most apparent in HDL where the top 10 highest ranked SNPs by RATE, RF, GBM, and BART were primarily located within two genes on the first and X chromosomes; but, the top 10 highest ranked SNPs by GOALS included 13 relevant genes across five different chromosomes (see Figures \ref{Fig4} \red{and \ref{Fig_S1}}). In addition to moderate signal on first and X chromosomes, GOALS also found signal on chromosomes 3, 11, 12,15, and 17. We hypothesize that GOALS prioritizes these additional genetic variants because it measures variable importance in function space and, thus, is better positioned to identify \red{features} whose associations are driven primarily by non-additive effects.} Importantly, many of the candidate SNPs selected by GOALS (and their respective genes) have been previously discovered by past publications as having some functional relationship with HDL content. For example, \textit{Tcq12}, \textit{Hyplip2}, and \textit{Ddiab41} have all been shown to associated with fat, cholesterol, and metabolism \citep{Gu:1999aa, valdar_gwas_2006, Moen:2007aa, Lawson:2011aa, Ostergren:2015aa, bult_mouse_2019}. 

There was notable overlap in the findings \red{for all methods} in the analysis of body weight and the percentage of CD8+ cells. For example, \red{each approach} identified strong signal on the X chromosome for body weight, a genomic region that was also validated by \citet{valdar_gwas_2006} in the original study (see Figures \ref{Fig_S2} \red{and \ref{Fig_S3}}). Here, all methods other than the random forest detected several adiposity-related genes, including \textit{Obq6} \citep{taylor_gender-influenced_1999, chen_number-of-X_2012} and \textit{Dbts2} \citep{Cheverud:2004aa}. For the percentage of CD8+ cells, \red{all methods identified} many genes on chromosome 17 which are known to greatly determine the ratio of T-cells \citep{yalcin_commercially_2010}, and some have been suggested to modulate cell adhesion and motility in the immune system \citep{Kim:2006aa} (see Figures \ref{Fig_S4} \red{and \ref{Fig_S5}}). Overall, out of the top ten most prioritized variables ranked by \red{\textit{post-hoc} approaches} GOALS and RATE, there was a 30\% overlap for body weight and a 90\% overlap for the percentage of CD8+ cells. For the latter, only GOALS prioritized a SNP on Chromosome 4 which harbors genes \textit{Lpq1} involved in pairwise interactions that are associated with lymphocyte percentage \citep{Miller:2020aa}.

\begin{figure}[ht]
\centering
\includegraphics[width = \textwidth]{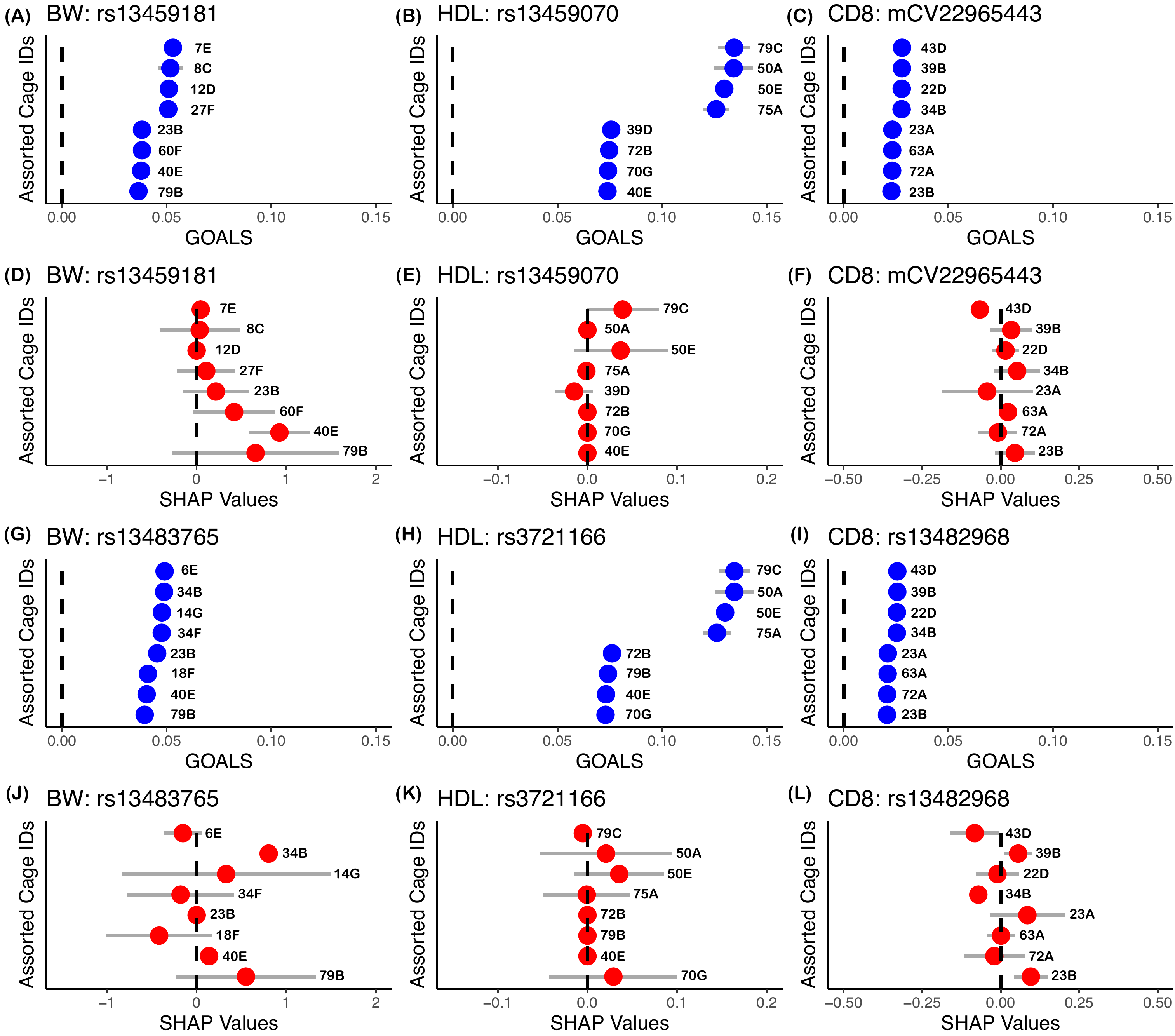}
\caption{\textbf{Plot of local variable importance according to GOALS \red{and SHAP} as a function of cage for notable genetic variants in the analysis of the heterogenous stock of mice data set from the Wellcome Trust Centre of Human Genetics \citep{valdar_simulating_2006, valdar_gwas_2006}.} Here, we show how SNPs have varying levels of importance for individual mice depending on the cage they were assigned to in the study. The traits analyzed here include \textbf{(A, D, G, J)} body weight (BW); \textbf{(B, E, H, K)} high-density lipoprotein (HDL) content; and \textbf{(C, F, I, L)} percentage of CD8+ cells. \red{The blue and red points are the mean local GOALS and SHAP values for each cage, respectively, and the grey lines show the total distribution for each score.} In this plot, we take the two SNPs with the greatest global GOALS value in each trait and plot the local values for the 4 cages with the greatest and least local means. The black dashed line is drawn at zero to represent a threshold where a SNP has no effect for a given mouse. \red{Note that due to computational considerations, SHAP is implemented by only considering all possible subsets of features on a given chromosome when computing local variable importance. Specifically, to run SHAP, we limit the data to include 372 SNPs on the X chromosome for body weight, 375 SNPs on chromosome 17 for percentage of CD8+ cells, and 758 SNPs on chromosome 3 for HDL content. GOALS is implemented on the full genome-wide data set.}}
\label{Fig5}
\end{figure}

Once again, the additional benefit of GOALS is its ability to also perform local variable importance for individual samples. In this particular data set from the Wellcome Trust Centre of Human Genetics, shared common environments between mice have been shown to contribute to the phenotypic variation of complex traits \citep{valdar_simulating_2006, valdar_gwas_2006,crawford_bayesian_2018}. For example, dietary and immunological phenotypes could depend heavily on the distribution of food and water in each cage. To that end, we assessed the local GOALS metrics for notable SNPs across mice according to the cages in which they were assigned during the study. In Figure \ref{Fig5}, we take the two SNPs with the greatest global GOALS value in each trait and plot the local values for the 4 cages with the greatest and least local means. \red{As a direct comparison, we also implement SHAP on the same set of SNPs. Note that due to computational considerations, SHAP is implemented by only considering all possible subsets of features on a given chromosome when computing local variable importance. Specifically, to run SHAP, we limit the data to include 372 SNPs on the X chromosome for body weight (runtime approximately 8 hours), 375 SNPs on chromosome 17 for percentage of CD8+ cells (runtime approximately 8 hours), and 758 SNPs on chromosome 3 for HDL content (runtime approximately 23 hours). GOALS, on the other hand, is implemented on the full genome-wide data set.} Overall, our proposed measure does indeed seem to capture environmental variation. This is again most apparent in HDL where the top SNPs rs13459070 (chromosome 3) and rs3721166 (chromosome 15) have very different effects on mice in different cages (e.g., Figure \ref{Fig5} \red{in panels B, E, H, and K}). Altogether, these sets of results would allow practitioners to perform deeper and more nuanced downstream analyses of phenotypic behavior in different populations.


\section*{Discussion}

In this paper, we proposed the “GlObal And Local Score” (GOALS) operator: a general approach for regression models that assesses variable importance for features at both the local and global levels of data, simultaneously. While this novel \textit{post hoc} interpretability measure can be used for any type of statistical model, we described the probabilistic properties of GOALS assuming we have fit a Gaussian process regression model with a shift-invariant covariance function. Through extensive simulations, we showed that our new measure can be used for feature selection and gives comparable state-of-the-art performance even in the presence of population structure (Figures \ref{Fig1}-\ref{Fig3}). The added benefit of GOALS is its ability to also understand how features affect individual samples on the local level and its computational efficiency to reach conclusions with much improved runtime as the dimensions of data increase. In applications to a real data set from the Wellcome Trust Centre of Human Genetics \citep{valdar_simulating_2006, valdar_gwas_2006}, we showed that GOALS has the ability to identify a greater number of trait-relevant genomic loci in a heterogenous stock of mice that have also been detected in many previous publications (Figures \ref{Fig4} and \red{\ref{Fig_S1}-\ref{Fig_S5}} and Tables \ref{Tab_S1}-\ref{Tab_S3}). The first key part of this analysis showed that GOALS incorporates non-additive information to find genetic signal that were missed by other approaches. The main takeaway from the real data analysis was that GOALS \red{can provide} local interpretability \red{which} enables downstream analyses to investigate how and why specific biomarkers are enriched for specific subsets of a population (Figure \ref{Fig5}). In this study, we saw how genetic variants associated with high density lipoprotein (HDL) content had varying local effects among mice assigned to different cages --- potentially, as a result of differing environments such as access to food and water. Ultimately, we hope that GOALS will encourage the continued development of probabilistic machine learning methods that can analyze complex data at the local and global levels.

The current implementation of the GOALS framework offers many directions for future development. First, while GOALS provides a measure of general association for nonlinear methods, it cannot be used to directly identify the component (i.e., linear versus nonlinear) that drives individual variable importance. Thus, despite being able to detect \red{variables} that are \red{important} to a response in a nonlinear fashion, GOALS is unable to directly identify the detailed orders of interaction effects. This same limitation also exists with distributional centrality measure such as RATE. A key part of our future work is to continue learning how to disentangle this information \citep[e.g., very similar to the goals of][]{woody_model_2021, kowal_fast_2021}. \red{As another extension}, GOALS does not enforce any sparsity or shrinkage when performing variable importance. Thus, while it has a natural null hypothesis (i.e., Eq.~\eq{GOALS}), \red{we do not provide} a significance threshold for variable selection. Common examples in \red{the statistics literature} include a Bonferroni-corrected threshold \citep{Gordon:2007aa} or selection based on a median probability model \citep{Barbieri:2004aa}. \red{One natural solution would be to utilize the posterior variances of $\bdelta^{(j)}$ and $\bar{\delta}^{(j)}$ which are derived and provided in the Supplementary Text. Another} natural solution could be to permute the \red{response variable} and refit the model a number of times to choose a GOALS-specific family-wise error rate (FWER) \citep[e.g.,][]{Hoti2006bayesianmapping,Stephens2009bayesianmethods}; however, this can be computationally intensive. One alternative could be to sample a collection of $\bdelta^{(j)}$ from the posterior distribution as specified in Eq.~\eq{GOALS_dist} and select significant variables based on a metric like a local false sign rate \citep{stephens_false_2016}. Lastly, univariate variable importance methods have been shown to be underpowered in settings where there are many causal variables with small effects. In many applications, \red{particularly in biomedicine,} recent methods have utilized prior knowledge to test groups of \red{variables} at a time to \red{improve power} \citep{liu2010versatile,Wu:2010aa,Carbonetto:2013aa,Leeuw:2015aa,Lamparter:2016aa,Nakka:2016aa,Zhu:2018aa,Sun:2019aa,Cheng:2020aa,Demetci:2021aa,ish2019interpreting}. This same group hypothesis extension can also be extended to the GOALS framework by simply perturbing multiple variables at a time.


\section*{Software Details}

Code for implementing the ``GlObal And Local Score'' (GOALS) operator is freely available at \url{https://github.com/lcrawlab/GOALS}, and is written in a combination of \texttt{R} and \texttt{C++} commands. \red{Software for computing the ``RelATive cEntrality'' (RATE) measure is carried out in \texttt{R} and \texttt{Python} code which is freely available at \url{https://github.com/lorinanthony/RATE}. The LASSO and elastic net regression models were run using the \texttt{glmnet} package in \texttt{R} \citep{Friedman:2010aa}, while SCANONE was implemented using the baseline \texttt{lm()} function in \texttt{R}. The random forest was fit using the \texttt{randomForest} package \citep{Liaw:2002aa}, the gradient boosting machine was fit using the \texttt{gbm} package \citep{Friedman:2001aa}, and the Bayesian additive regression tree was fit using the \texttt{BART} package \citep{Sparapani:2021aa} --- also all in \texttt{R}. Lastly, the SHapley Additive exPlanation approach was implemented using the \texttt{shap} package in \texttt{Python}.}


\section*{Acknowledgements}

This research was conducted using computational resources and services at the Center for Computation and Visualization (CCV), Brown University. E.T.~Winn-Nu\~{n}ez was supported by the National Science Foundation Graduate Research Program under Grant No.~1644760. This research was supported by a David \& Lucile Packard Fellowship for Science and Engineering awarded to L.~Crawford. \red{This study makes use of data generated by the Wellcome Trust Case Control Consortium (WTCCC). A full list of the investigators who contributed to the generation of the data is available from \url{www.wtccc.org.uk}. Funding for the WTCCC project was provided by the Wellcome Trust under award 076113, 085475, and 090355.} Any opinions, findings, and conclusions or recommendations expressed in this material are those of the author(s) and do not necessarily reflect the views of any of the funders.


\section*{Author Contributions}

All authors conceived the study and developed the methods. MG and LC supervised the project and provided resources. ETWN and LC developed the software. ETWN performed the analyses. All authors wrote and revised the manuscript.


\section*{Competing Interests}

The authors declare no competing interests.

 
\newpage 
\section*{Figures} 

\clearpage
\newpage
\bibliography{references}


\clearpage
\newpage

\setcounter{figure}{0}
\setcounter{table}{0}
\setcounter{equation}{0}
\makeatletter 
\renewcommand{\thefigure}{S\@arabic\c@figure} 
\makeatletter 
\renewcommand{\thetable}{S\@arabic\c@table} 
\renewcommand{\theequation}{S\@arabic\c@equation} 

\section*{Supplementary Text}

\subsection*{\red{Posterior Distribution for the Global Importance Scores in the GOALS}}

\red{In this subsection, we derive the full joint distribution for the sample means of the GOALS operator to conduct posterior inference on the global importance of features in a nonlinear model. As was done in the main text, consider a data set with $N$ individuals. We have an $N$-dimensional vector response variable $\by$ and an $N\times J$ design matrix $\bX$ with $J$ denoting the number of features. For consistency, we will demonstrate the properties of GOALS using a weight-space Gaussian process regression model 
\begin{align}
\by = \bm{f} + \bvarepsilon, \quad \quad \bm{f} \sim\N(\bm{0},\bK), \quad \quad \bvarepsilon \sim\N(\bm{0},\sigma^2\bI)
\end{align}
where $\bm{f} = [f(\bx_1),\ldots,f(\bx_N)]$ is an $N$-dimensional normally distributed random variable with mean vector $\bm{0}$ and a covariance matrix $\bK$ defined by some nonlinear kernel function.}

\red{In the main text, we defined a set of perturbed \red{features} $\bX+\bXi^{(j)}$, where $\bXi^{(j)}$ is an $N\times J$ matrix with rows $\bxi^{(j)}$ equal to all zeros except for the $j$-th element which we set to be a vector of some positive constant $\xi$. We then defined an $N$-dimensional vector $\bg^{(j)} = [f(\bx_1 + \bxi^{(j)}),\ldots,f(\bx_N + \bxi^{(j)})]$ where we showed that the joint distribution for the GOALS operator $\bdelta^{(j)} = \bm{f} - \bg^{(j)}$ (conditional on the data) can be written as the following
\begin{align}
\begin{bmatrix} \bdelta^{(1)} \\ \vdots \\ \bdelta^{(J)}\end{bmatrix} \bigg|\, \by \sim \N\left(\begin{bmatrix} \left[\bK - \big(\bB^{(1)}\big)^{\T}\right]\bA^{-1}\by \\ \vdots \\ \left[\bK - \big(\bB^{(J)}\big)^{\T}\right]\bA^{-1}\by \end{bmatrix}, \begin{bmatrix} \bSigma^{(1)} & \cdots & \bSigma^{(1,J)} \\ \vdots & \ddots & \vdots \\ \bSigma^{(J,1)} & \cdots & \bSigma^{(J)} \end{bmatrix} \right)
\end{align}
where, in addition to previous notation, $\bA = \bK + \sigma^2\bI$ is the marginal variance of the response vector $\by$; $\bB^{(j)}$ is the covariance between $\bm{f}$ and $\bg^{(j)}$ using the original matrix $\bX$ and the perturbed matrix $\bX+\bXi^{(j)}$; $\bC^{(j)}$ is the variance of $\bg^{(j)}$ using the perturbed matrix $\bX+\bXi^{(j)}$; and $\bD^{(j,l)}$ is the covariance between $\bg^{(j)}$ and $\bg^{(l)}$ having perturbed the $j$-th and $l$-th feature, respectively. Furthermore, we define
\begin{align*}
\bSigma^{(j)} &= \bK\bA^{-1}\bK - \big(\bB^{(j)}\big)^{\T}\bA^{-1}\bB^{(j)}-\left[\big(\bB^{(j)}\big)^{\T}-\big(\bB^{(j)}\big)^{\T}\bA^{-1}\bK+\bB^{(j)}-\bK\bA^{-1}\bB^{(j)}\right]\\
\bSigma^{(j,l)} &= \bK - \bK\bA^{-1}\bK + \bD^{(j,l)} - \big(\bB^{(j)}\big)^{\T}\bA^{-1}\bB^{(l)} - \left[\big(\bB^{(j)}\big)^{\T} - \big(\bB^{(j)}\big)^{\T}\bA^{-1}\bK+\bB^{(l)}-\bK\bA^{-1}\bB^{(l)} \right].
\end{align*}
Altogether, the above represents a joint conditional distribution from which one can sample estimates of each $\bdelta^{(j)}$ and obtain local interpretability. To investigate the global interpretability of each \red{feature}, one can use the sample mean across the local explanations for all observations where $\bar{\delta}^{(j)} = \bm{1}^{\T}\bdelta^{(j)}/N$ with $\bm{1}$ being an $N$-dimensional vector of ones. These global interpretability scores have the following joint distribution
\begin{align}
\begin{bmatrix} \bar\delta^{(1)} \\ \vdots \\ \bar\delta^{(J)}\end{bmatrix} \bigg|\, \by \sim \N\left(\begin{bmatrix} \bm{1}^{\T}\left[\bK - \big(\bB^{(1)}\big)^{\T}\right]\bA^{-1}\by/N \\ \vdots \\ \bm{1}^{\T}\left[\bK - \big(\bB^{(J)}\big)^{\T}\right]\bA^{-1}\by/N \end{bmatrix}, \begin{bmatrix} \bm{1}^{\T}\bSigma^{(1)}\bm{1}/N^2 & \cdots & \bm{1}^{\T}\bSigma^{(1,J)}\bm{1}/N^2 \\ \vdots & \ddots & \vdots \\ \bm{1}^{\T}\bSigma^{(J,1)}\bm{1}/N^2 & \cdots & \bm{1}^{\T}\bSigma^{(J)}\bm{1}/N^2 \end{bmatrix} \right).
\end{align}
Therefore, to simulate from the posterior distribution of the sample means, one simply needs to compute the following closed form equations for the first and second moments
\begin{align}
\begin{aligned}
\E\left[\bar\delta^{(j)}\right] &= \bm{1}^{\T}\left[\bK - \big(\bB^{(j)}\big)^{\T}\right]\bA^{-1}\by/N\\
\V\left[\bar\delta^{(j)}\right] &= \left(\lambda + \alpha_{jj} - 2\psi_j\right)/N^2\\
\V\left[\bar\delta^{(j)},\bar\delta^{(l)}\right] &= \left(\lambda + \alpha_{jl} - \psi_j-\psi_l\right)/N^2\\
\end{aligned}
\end{align}
where $\lambda = \bm{1}^{\T}\bK\bm{1}-\bm{1}^{\T}\bK\bA^{-1}\bK\bm{1}$; $\alpha_{jl} = \bm{1}^{\T}\bD^{(j,l)}\bm{1}-\bm{1}^{\T}\big(\bB^{(j)}\big)^{\T}\bA^{-1}\bB^{(l)}\bm{1}$; and $\psi_j = \bm{1}^{\T}\bB^{(j)}\bm{1} -\bm{1}^{\T}\bK\bA^{-1}\bB^{(j)}\bm{1}$, respectively.} 

\subsection*{\red{Extension of the GOALS to Probabilistic Neural Networks}}

\red{In this section, we show how the “GlObal And Local Score” (GOALS) operator can be used to determine global and local interpretability in probabilistic neural networks. In contrast to  a ``standard'' neural network, which uses maximum likelihood point-estimates for its parameters, we will assume a model architecture that places a prior distribution over its weights. During training, we will use a learned posterior probability over these weights to compute the posterior predictive distribution. Once again, we consider a general data application where we are given with an $N$-dimensional set of response variables $\by$ and an $N\times J$ design matrix $\bX$ with $J$ covariates. For this problem, we assume the following hierarchical network architecture to learn the predicted response in the data
\begin{align}
\by = r^{-1}(\bm{f}), \quad \quad \bm{f} = \bH(\bvartheta)\bw, \quad \quad \textbf{w} \sim \pi \, , \label{eq:logit-batch}
\end{align}
where $r(\bullet)$ is a link function (which we will assume to be the identity for regression-based tasks), $\bvartheta$ is a vector of inner layer weights, and $\bm{f}$ is an $N$-dimensional vector of smooth latent values or ``functions'' that need to be estimated. Here, we use $\bH(\bvartheta) = h(\bX\bvartheta)$ to denote an $N\times L$ matrix of activations from the penultimate layer (which are fixed given a predetermined activation function $h(\bullet)$, a set of \red{features} $\bX$, and point estimates for the inner layer weights $\bvartheta$), and $\bw\sim \pi$ is a $L$-dimensional vector of weights at the output layer assumed to follow prior distribution $\pi$. For simplicity, we omit the bias term in Eq.~\eq{eq:logit-batch} that is produced during the training phase. Also note that the structure of the hidden layers in the model above can be of any size or type, provided that we have access to draws of the posterior predictive distribution for the response variables.}

\red{The structure of Equation \eq{eq:logit-batch} is motivated by the fact that we are most interested in the posterior distribution of the latent variables $\bm{f}$. To this end, we follow previous work \citep{ish2019interpreting} and split the network architecture into three key components: \textit{(i)} an input layer of the original \red{features} $\bX$, \textit{(ii)} hidden layers $\bH(\bvartheta)$ where parameters are deterministically computed, and \textit{(iii)} the outer layer where the parameters and activations are treated as random variables. As the size of datasets in many application areas continues to grow, it has become common to train neural networks with algorithms that are based on variational Bayes and the stochastic optimization of a variational lower bound \citep{hinton1993keeping,barber1998ensemble,graves2011practical}. Here, the variational Bayes framework has the additional benefit of providing closed-form expressions for the posterior distribution of the weights in the outer layer $\bw$ and, subsequently, the functions $\bm{f}$.} 

\red{We will begin by first specifying a prior $\pi(\bw)$ over the weights and replace the intractable true posterior $p(\bw\cond\by) \propto p(\by \cond \bw) \pi(\bw)$ with an approximating family of distributions $q_{\bphi}(\bw)$ where $\bphi$ denotes a collection of free parameters. The overall goal of variational inference is to minimize the Kullback-Leibler divergence between the exact and approximate posterior distributions, respectively. This is equivalent to maximizing the so-called evidence lower bound where all parameters can be optimized jointly as follows
\begin{equation}
\underset{\bphi,\bvartheta}{\arg\max}\,\, \mathbb{E}_{q_{\bphi}(\bw)} \left[\log \, p(\by\cond\bw,\bvartheta) \right] -\eta\,\text{KL} (q_{\bphi}(\bw) \, \| \, \pi(\bw)).\label{eq:ELBO}
\end{equation}
Depending on the chosen variational family, the gradients of the minimized $\text{KL} (q_{\bphi}(\bw) \, \| \, \pi(\bw))$ may be available in closed-form, while gradients of the log-likelihood $\log\, p(\by\cond\bw, \bvartheta)$ are evaluated using Monte Carlo samples and the local reparameterization trick \citep{kingma2015variational}. Following this procedure, we obtain an optimal set of parameters for $q_{\bphi}(\bw)$, with which we can sample posterior draws for the outer layer. For simplicity, we will assume isotropic Gaussians as the family of approximating distributions
\begin{align}
q_{\bphi}(\bw) = \N(\bm{0},\bV), \label{eq:fac-q-phi}
\end{align}
where $\bm{0}$ is vector of zeros and $\bV$ is a diagonal covariance matrix. Using Equations \eq{eq:logit-batch} and \eq{eq:fac-q-phi}, we may derive the implied distribution over the latent function values using the affine transformation
\begin{align}
\bm{f} \sim \mathcal{N}(\bm{0}, \bH(\bvartheta)\bV\bH(\bvartheta)^{\T}).\label{eq:logit-posterior}
\end{align}
While the elements of $\bw$ are independent, dependencies in the input data (via the deterministic hidden activations $\bH(\bvartheta) = h(\bX\bvartheta)$) induce a non-diagonal covariance $\bK = \bH(\bvartheta)\bV\bH(\bvartheta)^{\T}$ between the elements of the latent function $\bm{f}$.} 

\red{Similar to what was shown with Gaussian process regression, to perform variable importance with the GOALS measure, we can define perturbed \red{features} $\bX+\bXi^{(j)}$, where $\bXi^{(j)}$ is an $N\times J$ matrix with rows $\bxi^{(j)}$ equal to all zeros except for the $j$-th element which we set to be a vector of some positive constant $\xi$, and we can also define an $N$-dimensional vector $\bg^{(j)} = [f(\bx_1 + \bxi^{(j)}),\ldots,f(\bx_N + \bxi^{(j)})]$. An analogous way to think about variable importance is to consider the expected change in the mean response given a $\xi$-unit increase in the corresponding covariate (holding all else constant). This again leads to the natural quantity to understand the importance of each variable by examining $\bdelta^{(j)} = \bm{f}-\bg^{(j)}$. Using Eq.~\eq{eq:logit-posterior}, the posterior mean of $\bdelta^{(j)}$ to perform local variable importance in neural networks also takes on the general form
\begin{align}
\E\left[\bdelta^{(j)}\cond\by\right] = \left[\bK - \big(\bB^{(j)}\big)^{\T}\right]\bA^{-1}\by.\label{S2}
\end{align}
There are a two main differences in this formulation when working with neural networks. First, the marginal variance $\bA = \bK + \sigma^2\bI$ can be estimated by using $\sigma^2\approx \V[\by - \bH(\bvartheta)\bw]$ which approximates the variance of residual training error in the penultimate layer \cite[e.g.,][]{Demetci:2021aa}. Second, we must find the covariance between $\bm{f}$ and $\bg^{(j)}$ using the original matrix $\bX$ and some perturbed matrix $\bX+\bXi^{(j)}$. To do so, note that using the perturbed matrix as the input to an already trained neural network (i.e., meaning model weights have already been estimated and frozen) allows us to directly estimate new hidden neurons $\bH^{(j)}(\bvartheta) = h[(\bX+\bXi^{(j)})\bvartheta]$. This implies that the covariance between $\bm{f}$ and $\bg^{(j)}$ can be written as a function of $\bH(\bvartheta)$ and $\bH^{(j)}(\bvartheta)^{\T}$, respectively, where $\bB^{(j)} = \bH(\bvartheta)\bV\bH^{(j)}(\bvartheta)^{\T}$. Lastly, as we did in the main text, one can take the sample means of the local importance values to get a measurement of global
importance.}

\subsection*{Scalable Computation for GOALS in Linear Regression}

In this section, we show that the “GlObal And Local Score” (GOALS) operator can also be efficiently computed in a linear regression framework. As \red{in the previous sections, we will assume that we have access to} an $N$-dimensional vector \red{response variable} $\by$ and an $N\times J$ \red{design} matrix $\bX$ with $J$ denoting the number of \red{features}. Next, consider a standard linear model 
\begin{align}
\by = \bm{f} + \bvarepsilon, \quad \quad \bm{f} = \bX\bbeta, \quad \quad \bvarepsilon \sim\N(\bm{0},\sigma^2\bI)\label{S1}
\end{align}
where the function to be estimated $\bm{f}$ is assumed to be a linear combination of \red{features} in $\bX$ and their respective effects denoted by the $J$-dimensional vector $\bbeta = (\beta_1,\ldots,\beta_J)$ additive coefficients, $\bvarepsilon$ is a normally distributed error term with mean zero and scaled variance term $\sigma^2$, and $\bI$ denotes an $N\times N$ identity matrix. For convenience, we will assume that the \red{outcome variable $\by$} has been mean-centered and standardized. The key identity in this section is that we can equivalently represent the regression in Eq.~\eq{S1} as a Gaussian process model with a linear gram kernel where the covariance matrix is written as $\bK = \bX\bX^{\T}$. Once again, we will work with the posterior mean of $\bdelta^{(j)}$ of the GOALS measure which again takes on the same general form presented in Eq.~\eq{S2}. Since we are working within the context of linear regression, the covariance between $\bm{f}$ and $\bg^{(j)}$ simplifies to the following
\begin{align}
\bB^{(j)} = k(\bX,\bX+\bXi^{(j)}) = \bX(\bX+\bXi^{(j)})^{\T} = \bK + \bX\bXi^{(j)\T}.\label{S3}
\end{align}
Note that, because $\bXi^{(j)\T}$ is a matrix of all zeros except for the $j$-th column, we can use Eq.~\eq{S3} to simplify the form of Eq.~\eq{S2} as the following
\begin{align}
\E\left[\bdelta^{(j)}\cond\by\right] = -\bX\bXi^{(j)\T}\bA^{-1}\by = \xi\bx_{\bullet j}\bm{1}^{\T}\bA^{-1}\by
\end{align}
where $\bm{1}$ is an $N$-dimensional vector of ones and $\bx_{\bullet j}$ is the $j$-th column in the design matrix $\bX$. The main summary is that the computation of Eq.~\eq{S2} only relies on linear operations after an initial pre-computation of the term $\bm{1}^{\T}\bA^{-1}\by$ which can be sped up using matrix decompositions.  


\clearpage


\begin{figure}[ht]
\centering
\vspace*{-0.9 in}
\hspace*{-0.3 in}
\includegraphics[scale = 0.7]{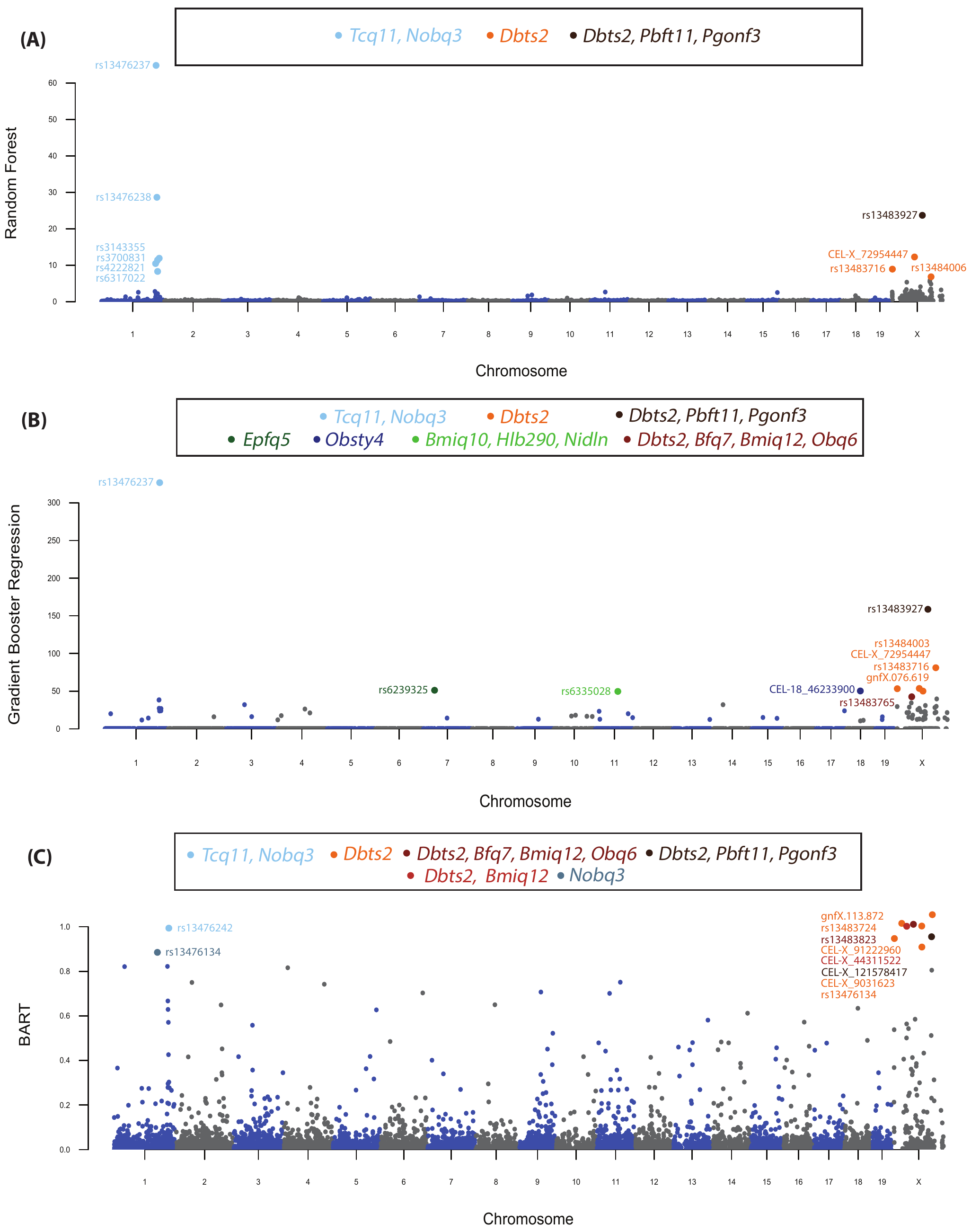}
\caption{\red{\textbf{Manhattan plot of variant-level association mapping results for high-density lipoprotein (HDL) content in the heterogenous stock of mice data set from the Wellcome Trust Centre of Human Genetics \citep{valdar_simulating_2006, valdar_gwas_2006} using competing global variable importance approaches.} Panel \textbf{(A)} depicts the global importance for each SNP plotted against their genomic positions after running a random forest (RF) with 500 trees \citep{Ishwaran:2019aa}. Here, genetic features are ranked by assessing their relative influence which is computed by taking the average total decrease in the residual sum of squares after splitting on each variable. As a direct comparison, we also include results after implementing \textbf{(B)} a gradient boosting machine (GBM) \citep{Friedman:2001aa} with 100 trees and \textbf{(C)} a Bayesian additive regression tree (BART) \citep{Chipman:2010aa} with 200 trees and 1000 MCMC iterations on the same quantitative trait. In the GBM, global importance is also determined by computing the relative influence of each SNP; while, in BART, features are ranked by the average number of times that they are used in decisions for each tree. In this figure, chromosomes are shown in alternating colors for clarity. The top 10 highest ranked SNPs by each method are labeled and color coded based on their nearest mapped gene(s) as cited by the Mouse Genome Informatics database (\url{http://www.informatics.jax.org/}) \citep{bult_mouse_2019}. These annotated genes are listed in the legends of each panel. A complete list of the values for all SNPs can be found in Table \ref{Tab_S2}.}}
\label{Fig_S1}
\end{figure}


\begin{landscape}

\begin{figure}[ht]
\centering
\vspace*{-0.7 in}
\hspace*{-0.3 in}
\includegraphics[scale = 0.65]{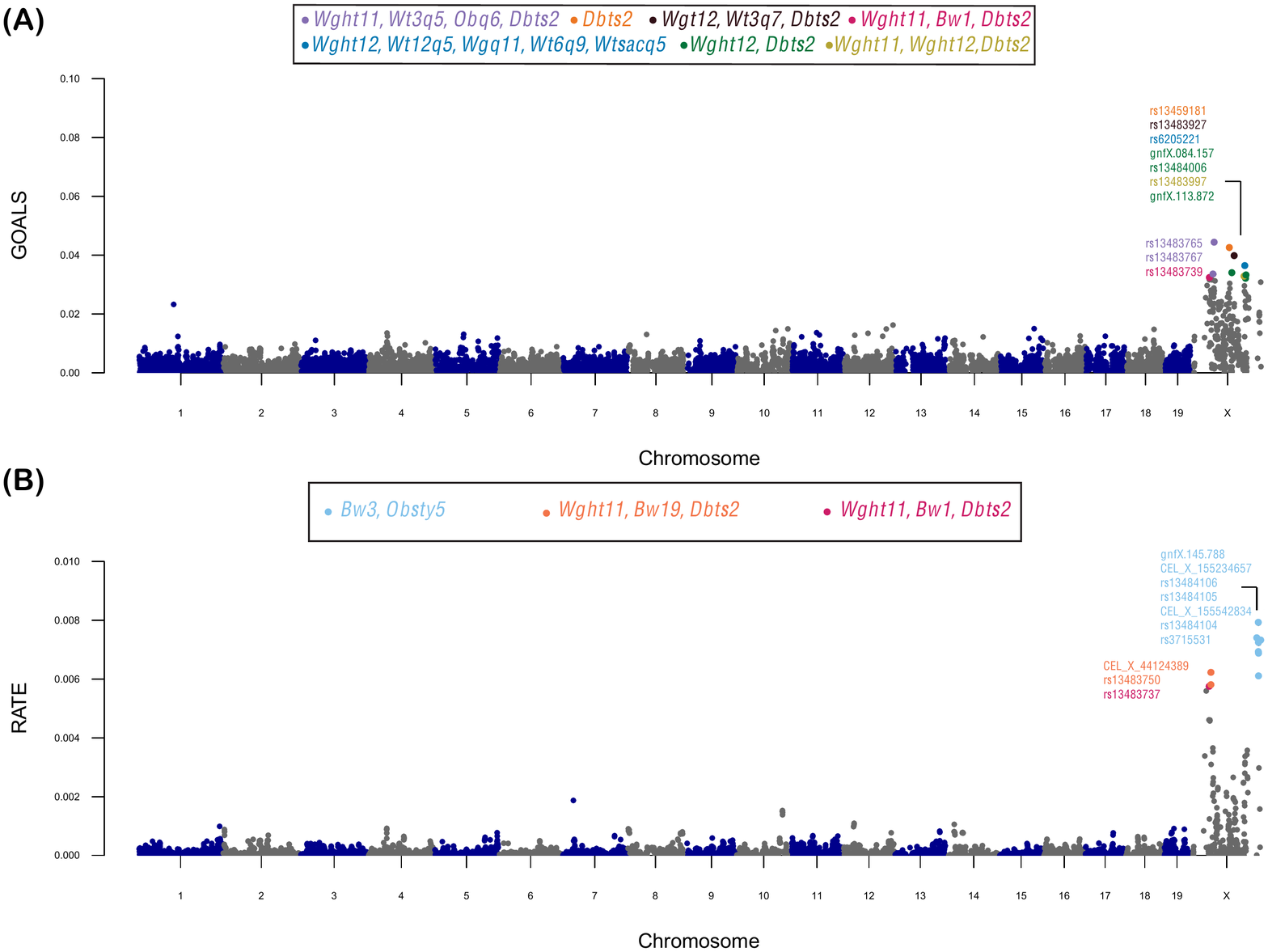}
\caption{\textbf{Manhattan plot of variant-level association mapping results for body weight in the heterogenous stock of mice data set from the Wellcome Trust Centre of Human Genetics \citep{valdar_simulating_2006, valdar_gwas_2006}.} Panel \textbf{(A)} depicts the global GOALS measure \red{(with $\xi = 1$)} of quality-control-positive SNPs plotted against their genomic positions after running a Bayesian Gaussian process (GP) regression on the quantitative trait. As a direct comparison, in panel \textbf{(B)}, we also include results after implementing RATE on the same fitted GP model. In this figure, chromosomes are shown in alternating colors for clarity. The top 10 highest ranked SNPs by GOALS and RATE, respectively, are labeled and color coded based on their nearest mapped gene(s) as cited by the Mouse Genome Informatics database (\url{http://www.informatics.jax.org/}) \citep{bult_mouse_2019}. These annotated genes are listed in the legends of each panel. A complete list of the GOALS and RATE values for all SNPs can be found in Table \ref{Tab_S2}.}
\label{Fig_S2}
\end{figure}

\end{landscape}


\begin{figure}[ht]
\centering
\vspace*{-0.9 in}
\hspace*{-0.3 in}
\includegraphics[scale = 0.6]{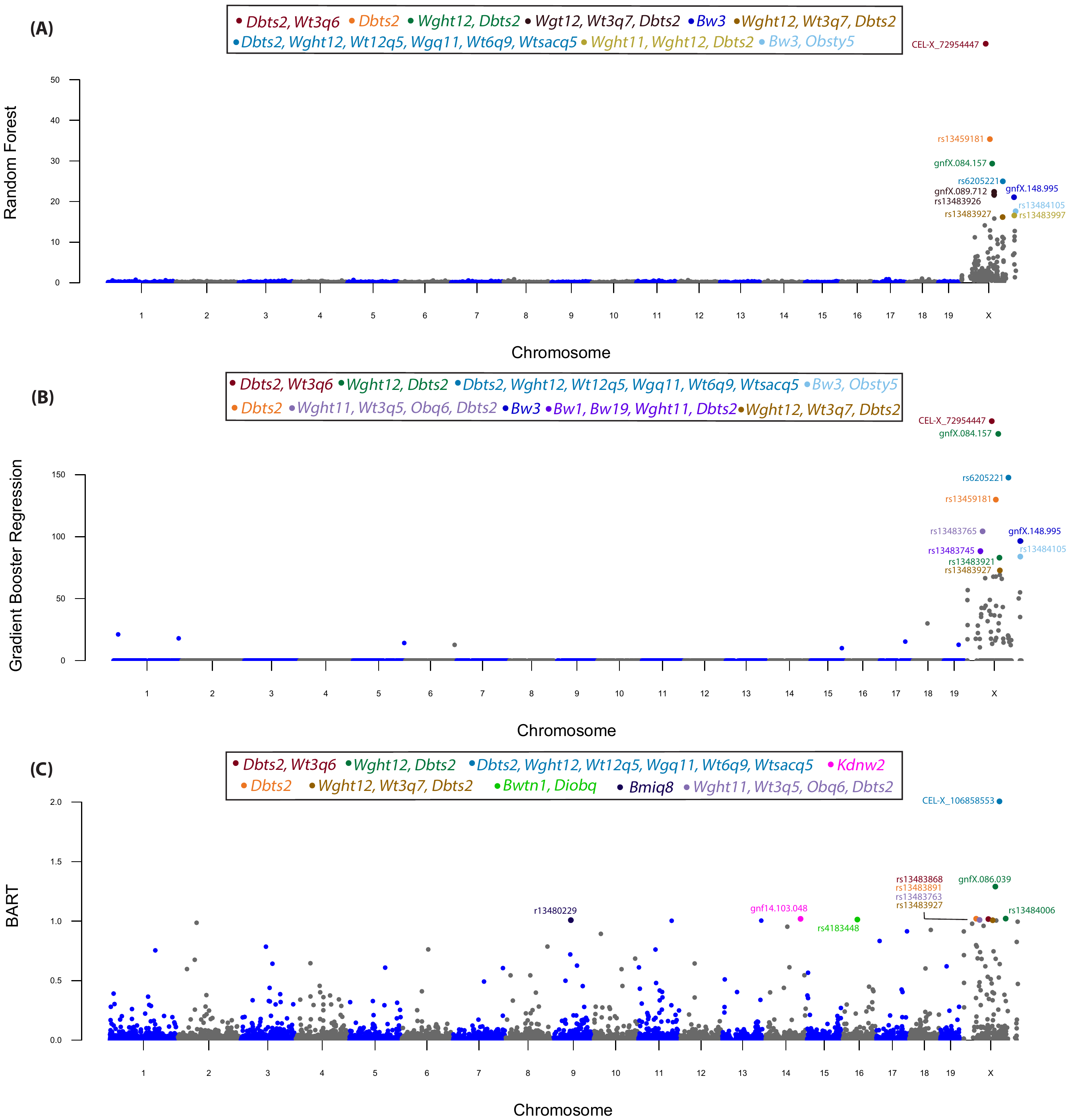}
\caption{\red{\textbf{Manhattan plot of variant-level association mapping results for body weight in the heterogenous stock of mice data set from the Wellcome Trust Centre of Human Genetics \citep{valdar_simulating_2006, valdar_gwas_2006} using competing global variable importance approaches.} Panel \textbf{(A)} depicts the global importance for each SNP plotted against their genomic positions after running a random forest (RF) with 500 trees \citep{Ishwaran:2019aa}. Here, genetic features are ranked by assessing their relative influence which is computed by taking the average total decrease in the residual sum of squares after splitting on each variable. As a direct comparison, we also include results after implementing \textbf{(B)} a gradient boosting machine (GBM) \citep{Friedman:2001aa} with 100 trees and \textbf{(C)} a Bayesian additive regression tree (BART) \citep{Chipman:2010aa} with 200 trees and 1000 MCMC iterations on the same quantitative trait. In the GBM, global importance is also determined by computing the relative influence of each SNP; while, in BART, features are ranked by the average number of times that they are used in decisions for each tree. In this figure, chromosomes are shown in alternating colors for clarity. The top 10 highest ranked SNPs by each method are labeled and color coded based on their nearest mapped gene(s) as cited by the Mouse Genome Informatics database (\url{http://www.informatics.jax.org/}) \citep{bult_mouse_2019}. These annotated genes are listed in the legends of each panel. A complete list of the values for all SNPs can be found in Table \ref{Tab_S2}.}}
\label{Fig_S3}
\end{figure}


\begin{landscape}

\begin{figure}[ht]
\centering
\vspace*{-0.7 in}
\hspace*{-0.3 in}
\includegraphics[scale = 0.55]{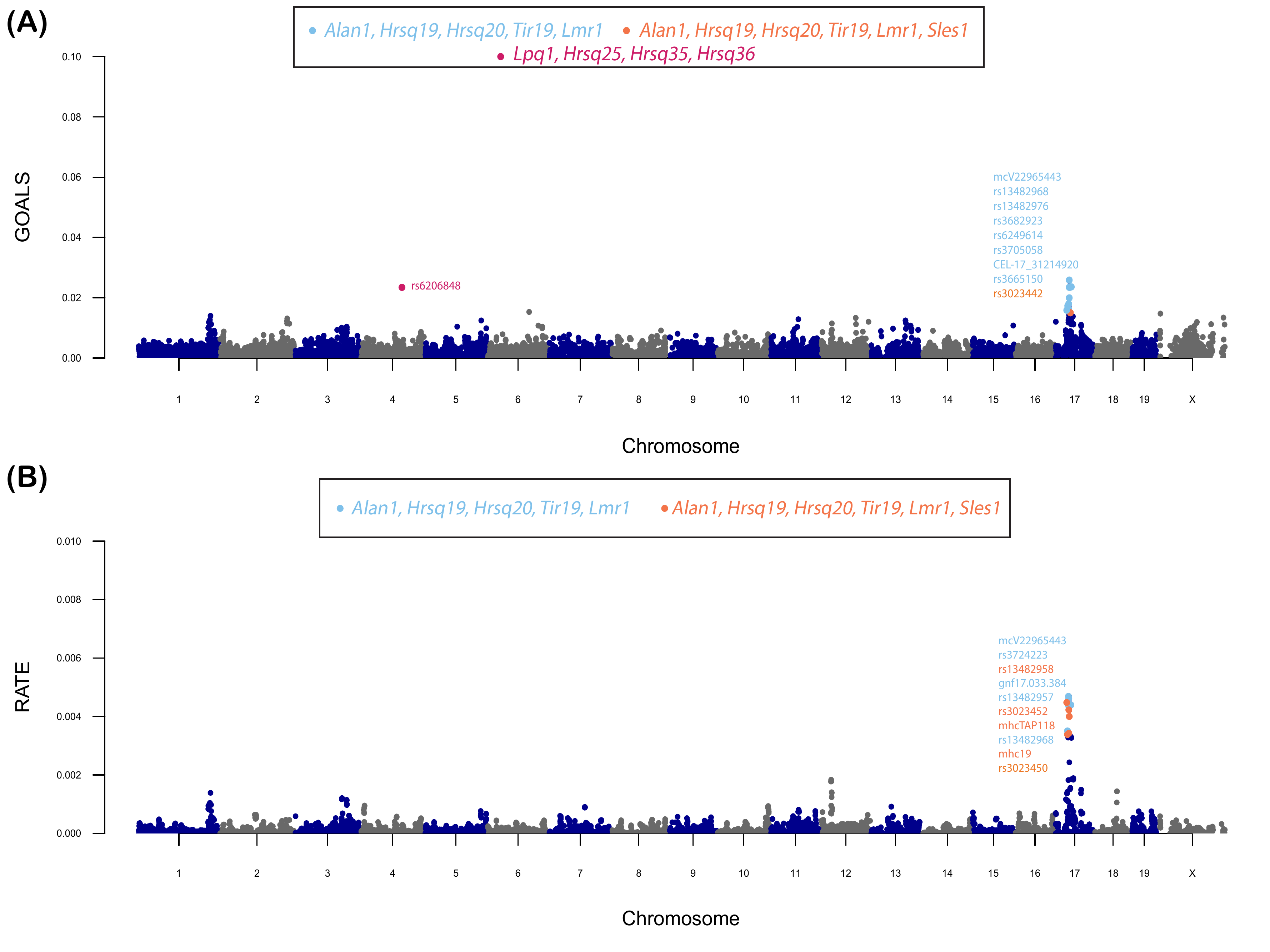}
\caption{\textbf{Manhattan plot of variant-level association mapping results for the percentage of CD8+ cells in the heterogenous stock of mice data set from the Wellcome Trust Centre of Human Genetics \citep{valdar_simulating_2006, valdar_gwas_2006}.} Panel \textbf{(A)} depicts the global GOALS measure \red{(with $\xi = 1$)} of quality-control-positive SNPs plotted against their genomic positions after running a Bayesian Gaussian process (GP) regression on the quantitative trait. As a direct comparison, in panel \textbf{(B)}, we also include results after implementing RATE on the same fitted GP model. In this figure, chromosomes are shown in alternating colors for clarity. The top 10 highest ranked SNPs by GOALS and RATE, respectively, are labeled and color coded based on their nearest mapped gene(s) as cited by the Mouse Genome Informatics database (\url{http://www.informatics.jax.org/}) \citep{bult_mouse_2019}. These annotated genes are listed in the legends of each panel. A complete list of the GOALS and RATE values for all SNPs can be found in Table \ref{Tab_S3}.}
\label{Fig_S4}
\end{figure}

\end{landscape}


\begin{figure}[ht]
\centering
\vspace*{-0.9 in}
\hspace*{-0.3 in}
\includegraphics[scale = 0.7]{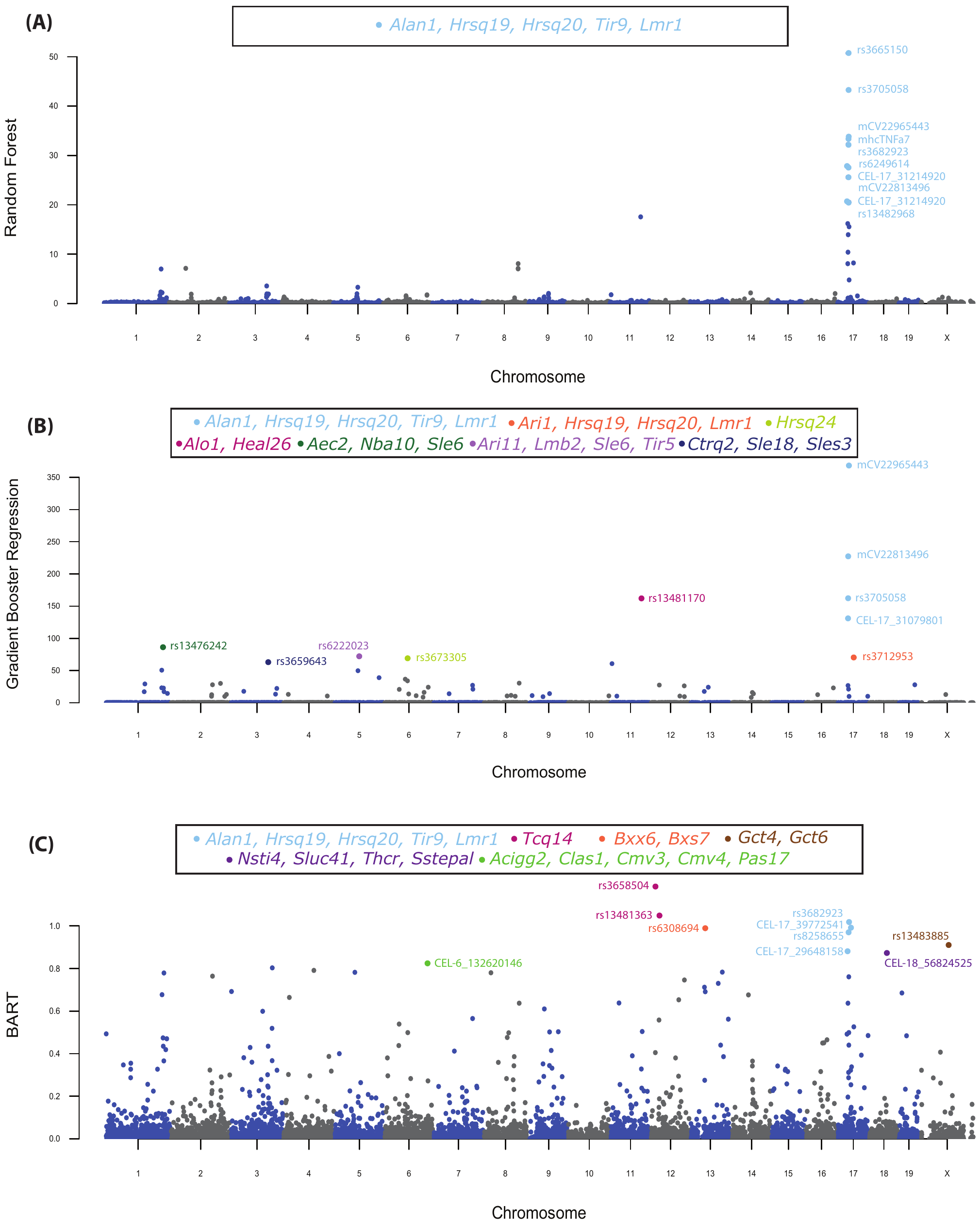}
\caption{\red{\textbf{Manhattan plot of variant-level association mapping results for the percentage of CD8+ cells in the heterogenous stock of mice data set from the Wellcome Trust Centre of Human Genetics \citep{valdar_simulating_2006, valdar_gwas_2006} using competing global variable importance approaches.} Panel \textbf{(A)} depicts the global importance for each SNP plotted against their genomic positions after running a random forest (RF) with 500 trees \citep{Ishwaran:2019aa}. Here, genetic features are ranked by assessing their relative influence which is computed by taking the average total decrease in the residual sum of squares after splitting on each variable. As a direct comparison, we also include results after implementing \textbf{(B)} a gradient boosting machine (GBM) \citep{Friedman:2001aa} with 100 trees and \textbf{(C)} a Bayesian additive regression tree (BART) \citep{Chipman:2010aa} with 200 trees and 1000 MCMC iterations on the same quantitative trait. In the GBM, global importance is also determined by computing the relative influence of each SNP; while, in BART, features are ranked by the average number of times that they are used in decisions for each tree. In this figure, chromosomes are shown in alternating colors for clarity. The top 10 highest ranked SNPs by each method are labeled and color coded based on their nearest mapped gene(s) as cited by the Mouse Genome Informatics database (\url{http://www.informatics.jax.org/}) \citep{bult_mouse_2019}. These annotated genes are listed in the legends of each panel. A complete list of the values for all SNPs can be found in Table \ref{Tab_S2}.}}
\label{Fig_S5}
\end{figure}


\clearpage
\newpage

\section*{Supplementary Tables}

\begin{table}[ht!]
\centering
\caption{\textbf{Genome-wide results for all SNPs in the heterogenous stock of mice data set while analyzing high-density lipoprotein (HDL).} Listed are the RATE and GOALS values for each SNP as computed via Gaussian Processes and the effect size analog. \red{As a direct comparison, we also include variable importance scores for each SNP after running a random forest (RF), a gradient boosting machine (GBM), and a Bayesian additive regression tree (BART).} Also listed are the chromosome location and physical position (bp) for each SNP. \red{This supplemental table can also be accessed on the Harvard Dataverse (\url{https://dataverse.harvard.edu/dataset.xhtml?persistentId=doi:10.7910/DVN/S6GUK4&faces-redirect=true)}.} (XLSX)}
\label{Tab_S1}
\end{table}

\begin{table}[ht!]
\centering
\caption{\textbf{Genome-wide results for all SNPs in the heterogenous stock of mice data set while analyzing body weight.} Listed are the RATE and GOALS values for each SNP as computed via Gaussian Processes and the effect size analog. \red{As a direct comparison, we also include variable importance scores for each SNP after running a random forest (RF), a gradient boosting machine (GBM), and a Bayesian additive regression tree (BART).} Also listed are the chromosome location and physical position (bp) for each SNP. \red{This supplemental table can also be accessed on the Harvard Dataverse (\url{https://dataverse.harvard.edu/dataset.xhtml?persistentId=doi:10.7910/DVN/S6GUK4&faces-redirect=true)}.} (XLSX)}
\label{Tab_S2}
\end{table}

\begin{table}[ht!]
\centering
\caption{\textbf{Genome-wide results for all SNPs in the heterogenous stock of mice data set while analyzing percentage of CD8+ cells.} Listed are the RATE and GOALS values for each SNP as computed via Gaussian Processes and the effect size analog. \red{As a direct comparison, we also include variable importance scores for each SNP after running a random forest (RF), a gradient boosting machine (GBM), and a Bayesian additive regression tree (BART).} Also listed are the chromosome location and physical position (bp) for each SNP. \red{This supplemental table can also be accessed on the Harvard Dataverse (\url{https://dataverse.harvard.edu/dataset.xhtml?persistentId=doi:10.7910/DVN/S6GUK4&faces-redirect=true)}.} (XLSX)}
\label{Tab_S3}
\end{table}

 
 \end{document}